\documentclass[twocolumn,twocolappendix]{aastex63} 
\usepackage{graphicx} 
\usepackage{gensymb}

\newcommand{\hi}{H\,{\sc i}\ }
\newcommand{\sersic}{S\'ersic }

\newcommand{\Mk}{ M$_{Ks}$ }
\newcommand{\MK}{ M$_{Ks}$ }
\newcommand{\mk}{ m$_{Ks}$ }
\newcommand{\tot}{355 }
\newcommand{\new}{264 }
\newcommand{\high}{134 }
\newcommand{\highlm}{36 }
\newcommand{\highsm}{98 }

\newcommand{\LMCLFer}{4.0$\pm$1.4 }
\newcommand{\SMCLFer}{2.1$\pm$0.6 }

\newcommand{\lowlimsm}{1.2$\pm$0.4 }

\newcommand{\lowlimlm}{1.6$\pm$0.7 }

\newcommand{\UA}{\affiliation{Steward Observatory, University of Arizona, 933 North Cherry Avenue, Tucson, AZ 85721-0065, USA}}

\newcommand{\Dart}{\affiliation{Department of Physics and Astronomy, Dartmouth College, 6127 Wilder Laboratory, Hanover, NH 03755, USA}}
	
\graphicspath{{figures/}}

\begin{document}

\title{Identifying Dwarfs of MC Analog GalaxiEs (ID-MAGE): The Search for Satellites Around Low-Mass Hosts}

\correspondingauthor{Laura Congreve Hunter}
\email{laura.c.hunter@dartmouth.edu}
\author[0000-0001-5368-3632]{Laura Congreve Hunter}
\Dart

\author[0000-0001-9649-4815]{\textsc{Bur\c{c}{\rlap{\.}\i}n Mutlu-Pakd{\rlap{\.}\i}l}}
\Dart

\author[0000-0003-4102-380X]{David~J. Sand}
\UA

\author[0000-0001-8354-7279]{Paul Bennet}
\affiliation{Space Telescope Science Institute, 3700 San Martin Drive, Baltimore, MD 21218, USA}

\author[0000-0002-7013-4392]{Donghyeon J. Khim}
\UA

\author[0000-0002-1763-4128]{Denija Crnojevi\'{c}}
\affil{Department of Physics \& Astronomy, University of Tampa, 401 West Kennedy Boulevard, Tampa, FL 33606, USA}

\author[0000-0001-9775-9029]{Amandine Doliva-Dolinsky}
\Dart
\affil{Department of Physics, University of Surrey, Guildford, Surrey GU2 7XH, UK}
\affil{Department of Physics \& Astronomy, University of Tampa, 401 West Kennedy Boulevard, Tampa, FL 33606, USA}

\author[0009-0004-9516-9593]{Emmanuel Durodola}
\Dart

\author[0000-0001-8245-779X]{Catherine Fielder}
\UA

\author{Rowan Goebel-Bain}
\Dart

\author[0000-0002-5434-4904]{Michael G. Jones}
\UA

\author[0000-0001-8855-3635]{Ananthan Karunakaran}
\affiliation{Department of Astronomy \& Astrophysics, University of Toronto, Toronto, ON M5S 3H4, Canada}

\author[0000-0002-0956-7949]{Kristine Spekkens}
\affiliation{Department of Physics, Engineering Physics and Astronomy, Queen’s University, Kingston, ON K7L 3N6, Canada}

\author[0000-0002-5177-727X]{Dennis Zaritsky}
\UA

\begin{abstract}
    We present results from ID-MAGE (Identifying Dwarfs of MC Analog GalaxiEs), a survey aimed at identifying and characterizing unresolved satellite galaxies around 35~nearby LMC- and SMC-mass hosts (D$=$4$-$10~Mpc).  We use archival DESI Legacy Survey imaging data and perform an extensive search for dwarf satellites, extending out to a radius of 150~kpc ($\sim$$R_{vir}$). We identify \tot candidate satellite galaxies, including \new new discoveries.  Extensive tests with injected galaxies demonstrate that the survey is complete down to $M_V\sim-$9.0 (assuming the distance of the host) and $\mu_{0,V}\sim$26 mag arcsec$^{-2}$ (assuming a n$=$1 \sersic profile). We perform consistent photometry, via \sersic profile fitting, on all candidates and have initiated a comprehensive follow-up campaign to confirm and characterize candidates. Through a systematic visual inspection campaign, we classify the top candidates as high-likelihood satellites. On average, we find \LMCLFer high-likelihood candidate satellites per LMC-mass host and \SMCLFer per SMC-mass host which is within the range predicted by cosmological models.  We use this sample to establish upper and lower estimates on the satellite luminosity function of LMC/SMC-mass galaxies. ID-MAGE nearly triples the number of low-mass galaxies surveyed for satellites with well-characterized completeness limits, providing a unique dataset to explore small-scale structure and dwarf galaxy evolution around low-mass hosts in diverse environments.

\end{abstract}

\section{Introduction}

Dwarf galaxies are unique laboratories for studying the nature of dark matter and galaxy formation. Because of their shallow gravitational potentials, they are extremely sensitive to differences in cosmological and galaxy formation and evolution models, resulting in a range of testable predictions for galaxy formation. Dwarfs can also be dramatically changed by their environment. Currently, our understanding of dwarfs is predominantly based on observations of satellite galaxies around Milky Way (MW)-mass galaxies \citep[e.g.,][]{Martin2013,Laevens15,Koposov2015,Drlica2015,Drlica2020,Chiboucas09,Chiboucas2013,Crnojevic16,Crnojevic2019,Muller2019,Smercina18,Sand14,toloba16,BMP22,BMP24,Bennet19,ELVES,SAGAI,SAGA}, which poses a risk of over-tailoring our models to observations  {of satellites systems of Milky Way-mass galaxies}. Therefore, a statistical sample of satellites of dwarf galaxies is urgently needed. 

One particular point of interest is lower-mass host systems, which are testing grounds to study the effect of environment on dwarf galaxy formation \citep{Wetzel2015}, where gravitational effects are reduced. The standard $\Lambda$ Cold Dark Matter ($\Lambda$CDM) model predicts that even moderate-sized dwarf galaxies should host their own small satellites \citep{Munshi2019}. The spatial clustering of $\sim$20 recently discovered ultra-faint ($M_{\star}$$<$$10^5 M_{\odot}$, $L$$<$$10^5 M_{\odot}$) galaxies provides evidence that the LMC itself might have fallen into the Local Group with its own satellite system \citep{Patel18,Battaglia22}. However, it also raises an interesting question about the luminosity function of the Magellanic association. The LMC has a fairly massive companion, the SMC, which is about 1.5 mag fainter than the LMC, but its next most luminous satellite is nearly 13~mag fainter (Hydrus~1, $M_V\sim-$4.7; \citealt{Koposov18}). This results in a $>$$10$~mag ‘gap’ in the satellite luminosity function.  This large magnitude-gap is unexpected given the number of substructures expected for a DM halo containing a galaxy as luminous as the LMC. From simulations, $\sim$$2$$-$$6$ satellites with $M_{\star}$$>$$10^5 M_{\odot}$ are expected around an LMC-mass galaxy (e.g., \citealt{Dooley17}). M33 appears to be similarly lacking in bright satellites with just two known likely satellites with luminosities of $L\simeq2-3$$\times$$10^4$$L_\odot$ \citep{Martin09,Collins24}. As M33 and the LMC are satellites of more massive hosts, their satellite systems may have been impacted by the MW/Andromeda.
To fully understand the satellite population of LMC-mass galaxies and to test our models for populating hosts with satellites in general, it is essential to study the satellite systems of nearby LMC/SMC-mass galaxies. Although the Local Group is the only place to detect the extremely low luminosity satellites that are being found near the LMC, we can build up a sample of satellites with $M_{\star}$$\gtrsim$$10^5 M_{\odot}$ by targeting nearby low-mass galaxies. 

Recent efforts, including surveys like MADCASH (Magellanic Analog Dwarf Companions and  Stellar Halos; \citealt{MADCASH}), DELVE-DEEP (the DEEP component of DECam Local Volume Exploration Survey (DELVE); \citealt{DELVE}), LBT-SONG (LBT Satellites of Nearby Galaxies Survey; \citealt{LBT-SONG}), { and ELVES-Dwarf (Exploration of Local VolumE Satellites Dwarf survey; \citealt{ELVES-Dwarf})}
have begun building up a sample of satellites around isolated dwarf galaxy hosts (\citealt{Sand15,sand24,Rich12,MADCASH,Carlin21,Carlin24,Davis2021,Davis2024,mcnanna24,ELVES-Dwarf}).  MADCASH, DELVE-DEEP and the nearby component of LBT-SONG focus on isolated hosts with stellar masses between $10^{8}-10^{10} M_\odot$. These programs focus on systems within $\lesssim$$4.5$~Mpc, where dwarf satellites can be investigated using resolved stars from ground-based observations. Within this range, it is possible to push the discovery frontier of dwarf galaxies to fainter magnitudes, enabling a more profound understanding of their characteristics through their resolved stellar populations. ID-MAGE's goal is to build \textit{a statistical sample} of satellites of dwarf galaxies  using integrated light searches to test dark matter and galaxy formation theories effectively.  Surveys such as the Exploration of Local VolumE Satellites Survey (ELVES) and Satellites Around Galactic Analogs (SAGA) have demonstrated the value of building statistical samples of dwarf satellites around MW-mass hosts.  ELVES utilized similar unresolved techniques to search for satellites of 30 MW-mass hosts out to a distance of 12~Mpc (\citealt{ELVES}), while SAGA used spectroscopy to confirm satellites around 101 MW-mass hosts with distances between 25 and 41~Mpc (\citealt{SAGAI,SAGA}). 

ID-MAGE aims to find unresolved satellites around 36 low-mass host galaxies with distances between 4$-$10~Mpc through a modified version of the well-established integrated light search algorithm \citep{Bennet17}. ID-MAGE nearly triples the number of low-mass galaxies surveyed for satellites with well-characterized detection limits. Our satellite sample will provide vital clues for our understanding of galaxy evolution physics at these scales, and serve as a ground of comparison to the results obtained for the LMC and MW-mass hosts, yielding new insights into how host properties can affect satellite dwarf evolution.  Additionally, with 36 low-mass hosts, we will reach sufficient statistical power to use their satellite populations as a test to the theoretical predictions \citep[e.g.,][]{Dooley17,Nadler2022,Jahn2022} to constrain the physics that shapes the relationship between dwarf baryonic properties and DM halo mass.


This paper presents the overview and initial results of the new survey ID-MAGE, including the candidate satellites identified, their photometric properties, and comparisons with cosmological predictions. 
In Section~\ref{survey}, we describe the goals of ID-MAGE and our host selections. Section~\ref{detect} details the detection algorithm and the survey's completeness. Section~\ref{galfit} details the photometry of the candidates.  In Section~\ref{discussion}, we compare our candidates to known Local Volume dwarf galaxies and compare the number of satellite candidates we detected to simulation predictions. Section~\ref{follow-up} goes into our ongoing campaign to follow-up our candidate satellites. Finally, we summarize our key results in Section~\ref{conclusion}.

\section{Survey Description}\label{survey}

A robust sample of satellite galaxies around hosts with a wide range of masses is required to test $\Lambda$CDM models and our understanding of galaxy formation and evolution.  The goal of ID-MAGE is to identify satellite galaxy systems around LMC/SMC-mass hosts in diverse environments. 
In this section, we present the selection of our hosts and their characteristics. We initially describe our host selection criteria in Section~\ref{host_select}. Following that, we discuss the host environment in Section~\ref{host_env}, where we describe the relative isolation of our hosts and highlight interesting environmental features. In Section \ref{data}, we summarize our use of the DESI Legacy Survey imaging data for ID-MAGE.

\subsection{Host Selection} \label{host_select}

\begin{figure*}[!th]
    \centering
    \includegraphics[width=\linewidth]{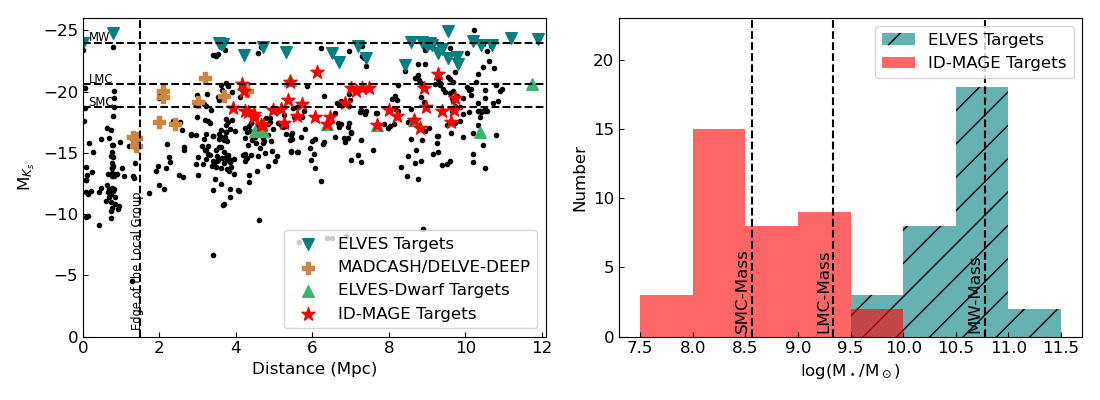}
    \vspace{-0.5cm}
    \caption{Left panel: Absolute Ks-band magnitudes (\Mk) versus distance for galaxies within 10~Mpc (black dots; \citealt{Kourkchi17} catalog).  The ID-MAGE hosts are shown in red stars.  For comparison, the ELVES host sample is shown in cyan triangles,  {ELVES-dwarf hosts' as green triangles,} and the MADCASH and DELVE-DEEP hosts are shown as brown pluses. The luminosities of the MW (\MK$=-24$; \citealt{BHG16}), LMC (\MK$=-20.6$; \citealt{Kourkchi17}), and SMC (\MK$=-18.6$; \citealt{Kourkchi17})  are shown as dashed lines.  Right panel: The stellar mass range of the ID-MAGE hosts (red), in comparison to the ELVES MW-mass host sample (cyan), assuming M$_*$/L$_K$ =0.6 \citep{McGaugh14}.  For reference, the masses of the MW (M$_*=6.08$$\times$$10^{10}$M$_\odot$; \citealt{Licquia15}), LMC (M$_*=2.2$$\times$$10^9$M$_\odot$), and SMC (M$_*=3.6$$\times$$10^8$M$_\odot$) derived from their K$_s$-band magnitude are marked with dashed black lines. ID-MAGE effectively extends the range of host masses surveyed. } 
    \label{fig:host_scatter}
\end{figure*}

We select our hosts from Cosmicflows-4 \citep{Tully23} --a catalog of the distances to 55,877 galaxies within z$=0.1$-- choosing galaxies simply via a cut in luminosity ($-17.0$ $\ge$ K$_s$ $\ge$ $-21.5$~mag), distance ($3.9$ $\le$ $D$ $\le$ $10$~Mpc), and Galactic latitude ($|b|$ $>$ $17^{\circ}$). Our magnitude cuts translate to a host stellar mass range of $\sim$10$^{8.0}$~M$_\odot$(one-fourth the SMC's stellar mass of M$_*=3.6$$\times$$10^8$M$_\odot$) to M$_*$$\sim$10$^{9.75}$ M$_\odot$ (three times the LMC's stellar mass of M$_*=2.2$$\times$$10^9$M$_\odot$), assuming M$_*$/L$_K$ =0.6 (\citealt{McGaugh14}). We require TRGB distances for our hosts to accurately match the distances of the systems with confidence.  We impose an absolute galactic latitude cut to avoid the plane of the MW. Additionally, we require coverage of the host galaxies out to a projected radius of 150~kpc in the DESI Legacy Imaging Surveys \citep{Dey19} Data Release (DR)-10 in the \textit{g-} and \textit{r-}bands. This ensures the full virial radius is covered by the imaging survey for each host. The hosts' viral radii are estimated as  {120$-$140~kpc} for LMC-mass galaxies and 80$-$100~kpc for SMC-mass galaxies \citep{Guo13,BMP2021}.  {Assuming \cite{Moster10}'s stellar mass-to-halo mass relationship, the halo masses in log(M$_{halo}$/M$_\odot$) of our LMC-mass galaxies ranges from 11.15 to 11.40 and 10.70 to 11.05 for our SMC-mass hosts.}  We exclude galaxies that fall within the projected virial radius of a MW-mass galaxy and have either $d \leq 1$~Mpc or $v \leq 100$kms$^{-1}$ relative to the host.  We also exclude host galaxies if there is a large known background galaxy cluster whose contamination would be difficult to remove from our satellite sample or if significant galactic cirrus impedes the search. In total, we exclude $\simeq$35 potential hosts within the Legacy Survey for the reasons described above. This results in a total of 36 low-mass galaxies (9 in the LMC-mass range, 27 in the SMC-mass range). The complete list of hosts is presented in Table~\ref{table:hosts_SMC}. 

Figure~\ref{fig:host_scatter} demonstrates how our survey expands the mass range of host galaxies surveyed for satellite systems, with clearly defined completeness limits. Our target host galaxies are considerably less massive than those in ELVES, aligning more closely with the mass ranges surveyed in programs like LBT-SONG \citep{LBT-SONG}, MADCASH \citep{MADCASH,Carlin24}, and DELVE-DEEP \citep{DELVE}.
Although most hosts are roughly LMC/SMC-like in stellar mass, our sample includes hosts both more and less massive than the LMC/SMC, which will allow insight into how host properties can affect satellite dwarf evolution. Unlike MADCASH and DELVE-DEEP, our selection criteria do not require the host galaxies to be isolated but they are required to not be satellites of larger galaxies (see Section~\ref{host_env}).

\begin{table*} 
    \footnotesize
    \centerwidetable
    \caption{ID-MAGE Host Properties}
        \begin{tabular}{l l l c c c c c c c r} 
        \hline
        \hline 
        Galaxy & RA & Dec & Dist & \mk & \Mk & m$_B$ & A$_B$ & M$_B$ & log(M$_*$/M$_\odot$) & $\Theta_5$ \\ 
         & J2000 & J2000 & Mpc & mag & mag & mag & \\
         (1) & (2) & (3) & (4) & (5) & (6) & (7) & (8) & (9) & (10) & (11) \\[0.5ex] 
        \hline 
        \multicolumn{11}{c}{LMC-Mass Hosts}\\ 
        \hline
        NGC~4449 & 12:28:11.2 & +44:05:40 & 4.16$\pm$0.02 & 7.49 & $-$20.61 & 9.99 & 0.08 & $-$18.19 & 9.33 & 0.4 \\
        NGC~4244 & 12:17:29.9 & +37:48:27 & 4.20$\pm$0.14 & 8.1 & $-$20.01 & 10.88 & 0.09 & $-$17.33. & 9.09 & 0.5 \\
        NGC~4605 & 12:40:00.3 & +61:36:29 & 5.41$\pm$0.05 & 7.92 & $-$20.75 & 10.89 & 0.06 & $-$17.84 & 9.38 & $-$0.6 \\
        NGC~6503 & 17:49:27.6 & +70:08:41 & 6.12$\pm$0.20 & 7.37 & $-$21.56 & 13.45 & 0.14 & $-$15.62 & 9.71 & $-$0.8 \\
        NGC~0672$^*$ & 01:47:53.2 & +27:26:01 & 7.00$\pm$0.26 & 8.98 & $-$20.25 & 11.31 & 0.34 & $-$18.26 & 9.18 & 3.8 \\
        NGC~0024 & 00:09:56.4 & $-$24:57:48 & 7.13$\pm$0.10 & 9.21 & $-$20.06 & 12.38 & 0.08 & $-$16.97 & 9.11 & $-$0.8 \\
        IC~1727$^*$ & 01:47:30.1 & +27:19:52 & 7.29$\pm$0.20 & 9.00 & $-$20.31 & 12.07 & 0.34 & $-$17.58 & 9.21 & 4.0 \\
        NGC~3432 & 10:52:31.1 & +36:37:08 & 8.9$\pm$0.80 & 9.43 & $-$20.32 & 11.67 & 0.06 & $-$18.14 & 9.21 & 3.3 \\
        NGC~7090 & 21:36:28.6 & $-$54:33:26 & 9.29$\pm$0.26 & 8.40 & $-$21.44 &  11.11 & 0.10 & $-$18.61 & 9.66 & $-$1.3 \\[1ex]
        \hline
        \multicolumn{11}{c}{SMC-Mass Hosts}\\
        \hline 
        NGC~0625 & 01:35:05.0 & $-$41:26:11 & 3.92$\pm$0.07 & 9.33 & $-$18.63  & 11.59 & 0.07 & $-$16.45 & 8.54 & $-$0.2 \\
        IC~4182 & 13:05:49.3 & +37:36:21 & 4.24$\pm$0.08 & 9.72 & $-$18.44 & 12.02 & 0.06 & $-$16.18 & 8.46 & 0.9 \\
        NGC~4236 & 12:16:43.3 & +69:27:56 & 4.31$\pm$0.08 & 9.82 & $-$18.35 & 10.03 & 0.06 & $-$18.18 & 8.43 & $-$0.1 \\
        ESO245-G05 & 01:45:03.6 & $-$43:35:53 & 4.46$\pm$0.12 & 10.3 & $-$17.95 & 12.7 & 0.07 & $-$15.62 & 8.26 & $-$0.5\\
        NGC~5204 & 13:29:36.4 & +58:25:04 & 4.48$\pm$0.50 & 10.12 & $-$18.14 & 11.73 & 0.05 & $-$16.58 & 8.34 & $-$0.4 \\
        NGC~4395 & 12:25:49.8 & +33:32:46 & 4.65$\pm$0.02 & 11.10 & $-$17.23 & 10.64 & 0.07 & $-$17.77 & 7.98 & 0.3 \\
        UGC~08201 & 13:06:24.9 & +67:42:25 & 4.72$\pm$0.04 & 11.10 & $-$17.27 & 13.31 & 0.10 & $-$15.16 & 7.99 & 0.1 \\
        ESO115-G21 & 02:37:40.7 & $-$61:21:06 & 4.96$\pm$0.05 & 10.00 & $-$16.81 & 13.34 & 0.11 & $-$15.25 & 8.48 & $-$1.0\\
        NGC~3738 & 11:35:48.6 & +54:31:22 & 5.19$\pm$0.05 & 10.00 & $-$18.57 & 12.12 & 0.05 & $-$16.51 & 8.51 & $-$0.4 \\
        NGC~0784 & 02:01:16.8 & +28:50:37 & 5.26$\pm$0.02 & 11.2 & $-$17.40 & 12.5 & 0.26 & $-$16.36 & 8.05 & $-$0.4 \\
        IC~5052 & 20:52:06.2 & $-$69:12:14 & 5.37$\pm$0.15 & 9.32 & $-$19.32 & 11.68 & 0.22 & $-$17.19 & 8.82 & $-$1.1 \\
        NGC~1705 & 04:54:13.5 & $-$53:21:39 & 5.61$\pm$0.10 & 10.76 & $-$17.98 & 12.77 & 0.03 & $-$16.00 & 8.28 & $-$1.4 \\
        ESO154-G23 & 02:56:50.4 & $-$54:34:23 & 5.74$\pm$0.05 & 9.80 & $-$18.99 & 12.71 & 0.07 & $-$16.15 & 8.68 & $-$1.1 \\
        IC~1959 & 03:33:11.8 & $-$50:24:38 & 6.07$\pm$0.11 & 11.03 & $-$17.89 & 13.2 & 0.05 & $-$15.77 & 8.24 & $-$1.1 \\
        NGC~4707 & 12:48:22.9 & +51:09:53 & 6.38$\pm$0.29 & 11.70 & $-$17.32 & 13.43 & 0.05 & $-$15.64 & 8.02 & $-$0.3 \\
        NGC~4455 & 12:28:44.1 & +22:49:21 & 6.46$\pm$0.27 & 11.11 & $-$17.94 & 12.93 & 0.09 & $-$16.21 & 8.26 & $-$0.4 \\
        NGC~5585 & 14:19:48.3 & +56:43:49 & 6.84$\pm$0.31 & 10.11 & $-$19.07 & 13.17 & 0.07 & $-$16.08 & 8.71 & $-$0.5 \\
        UGC~04115 & 07:57:01.8 & +14:23:17 & 7.70$\pm$0.11 & 12.10 & $-$17.33 & 15.23 & 0.12 & $-$14.32 & 8.02 & $-$1.2 \\
        UGC~03974 & 07:41:55.4 & +16:48:09 & 7.99$\pm$0.07 & 11.00 & $-$18.51 & 13.62 & 0.14 & $-$16.03 & 8.49 & 1.0 \\
        NGC~2188 & 06:10:09.5 & $-$34:06:22 & 8.22$\pm$0.23 & 11.6 & $-$17.97 & 12.14 & 0.14 & $-$17.57 & 8.27 & 0.9 \\
        UGC~05423 & 10:05:30.6 & +70:21:52 & 8.66$\pm$0.12 & 12.0 & $-$17.69 & 14.42 & 0.34 & $-$15.61 & 8.16 & $-$0.8\\
        ESO364-G29 & 06:05:45.4 & $-$33:04:54 & 8.81$\pm$0.33 & 12.73 & $-$16.99 & 13.67 & 0.19 & $-$16.24 & 7.88 & 0.1 \\
        IC~4951 & 20:09:31.8 & $-$61:51:02 & 9.0$\pm$0.6 & 11.00 & $-$18.77 & 13.97 & 0.02 & $-$15.81 & 8.59 & $-$0.8 \\
        HIPASSJ0607-34 & 06:01:19.7 & $-$34:12:16 & 9.4$\pm$0.4 & 11.50 & $-$18.37 & 14.09 & 0.15 & $-$15.93 & 8.43 & 1.6 \\
        UGC~04426 & 08:28:28.4 & +41:51:24 & 9.62$\pm$0.18 & 12.40 & $-$17.52 & 15.27 & 0.16 & $-$14.81 & 8.09 & $-$0.9 \\
        ESO486-G21 & 05:03:19.7 & $-$25:25:23 & 9.7$\pm$1.3 & 11.40 & $-$18.53 & 14.37 & 0.14 & $-$15.70 & 8.50 & 0.8 \\
        NGC~4861 & 12:59:02.0 & +34:51:37 & 9.71$\pm$0.18 & 10.50 & $-$19.44 & 12.9 & 0.04 & $-$17.09 & 8.86 & 0.4 \\[1ex]
        \hline 
    \end{tabular} 
    \tablecomments{Column~1: Galaxy Name. Column~2: the Right Ascension (J2000.0). Column~3: the Declination (J2000.0). Column~4: TRGB distances from \cite{EDD} in Mpc. Column~5: Apparent K$_s$-band magnitude. Column~6: Absolute K$_s$-band magnitude \citep{Kourkchi17}. Column~7: Apparent B-band magnitude. Column~8: B-band extinctions. Column~9: Absolute B-band magnitude \citep{Karachen13}. Column~10: The logarithm of the total stellar mass, which are determined using the K$_s$ luminosity and and M$_\star$/L$_{K_s}=$0.6, assuming M$^\odot_{K_s} = $3.27 in the Vega system \citep{Willmer18}. Column~11: Tidal Index from \cite{Karachen13}.\\    
    $\star$ NGC~0672 and IC~1727 are considered the same system with the search area covering both hosts' virial radii. 
    }
    \label{table:hosts_SMC} 
\end{table*}

In our sample, two LMC-mass galaxies, NGC~0672 and IC~1727,  
are located less than ten arcminutes apart in the sky ($\sim$16~kpc at 7.0~Mpc) and are nearly at the same distance (IC~1727 d$=$ 7.3 Mpc, NGC~672 d$=$7.0 Mpc). Because of their close proximity to each other, we treat them as a single group in our analysis, referring to it as the NGC0672/IC1727 system when discussing their candidate satellites. We conduct a dwarf search across the full virial radius of both galaxies and classify any identified satellites as associated with both galaxies.

\subsection{Host Environment} \label{host_env}

Environment plays a significant role in the evolution of satellite galaxies. The weaker tidal and ram pressure interactions associated with lower-mass hosts may enable their satellites to preserve their neutral gas reservoirs and continue forming stars (e.g., \citealt{spekkens14}). To explore how environmental factors affect satellite evolution, our selected host galaxies vary in their proximity to other nearby galaxies, ranging from complete isolation with no other galaxies within 500~kpc, to being part of groups of dwarf galaxies.

\begin{figure*}[!ht]

\includegraphics[width=\linewidth]{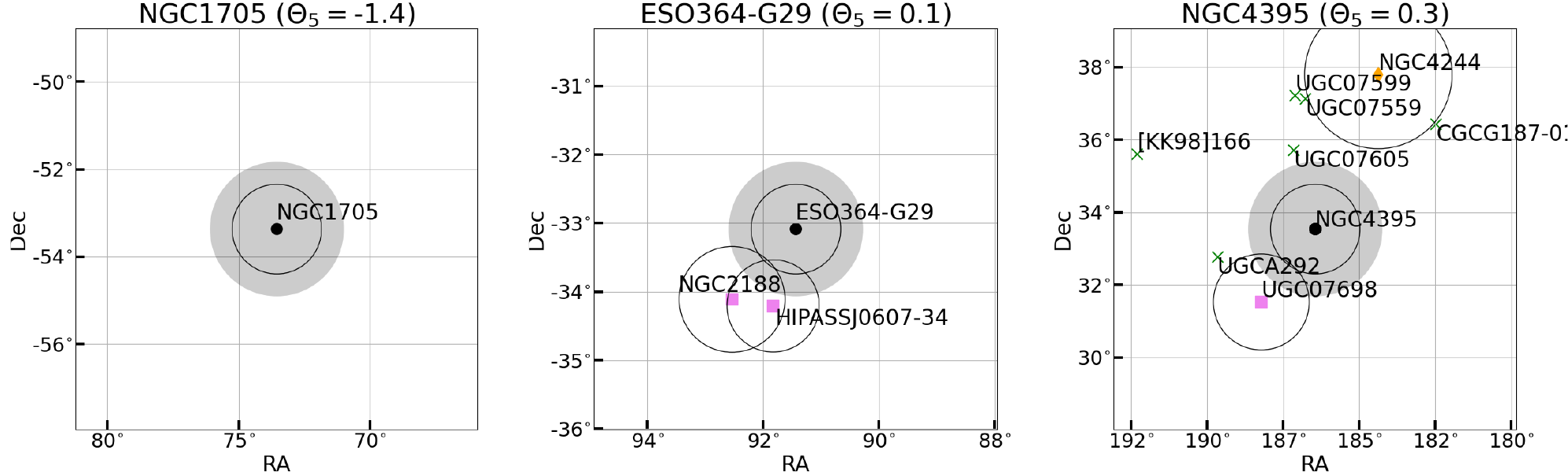}
\caption{The surrounding environment for three host systems in ID-MAGE, which can be very isolated (left) or in a more crowded environment (middle and right).  The central black point is the host galaxy, and the gray region is the 150 kpc radius search area.  The green crosses (x) are known galaxies that are less massive than our hosts (log(M$_*$)$<$7.5 M$\odot$), pink squares are SMC-mass galaxies, and the orange diamonds are LMC-mass galaxies. The black circles represent a rough estimate of the virial radius of each galaxy based on their K-band magnitudes reported in \cite{Kourkchi17}. For SMC-mass galaxies, the radius is 110~kpc; for LMC-mass galaxies, the radius is 150 kpc.  { The  figure set (35 images) of individual environment plots including identified satellites for each host  is available in the online journal.}}

\label{fig:tile_env}
\end{figure*}

We utilize the catalogs from Cosmicflows-4 \citep{EDD} and \cite{Kourkchi17} to check galaxies near our hosts. For each host, we select all galaxies within a projected radius of 450~kpc and with a distance in Cosmicflows-4 within 1~Mpc.  Figure~\ref{fig:tile_env} showcases examples of three environments in our host sample. The surveyed areas are marked in gray. Galaxies with \Mk$\leq -17$ are shown with their approximate virial radii indicated: 110~kpc for SMC-mass galaxies, 150~kpc for LMC-mass galaxies, and 300~kpc for more massive galaxies. As shown in Figure~\ref{fig:tile_env}, our surveyed hosts are in diverse environments, ranging from those with no known galaxies with \Mk$\leq -17$ within a projected radius of 450~kpc at a similar distance (left panel), to hosts that are part of larger galaxy groups (middle and right panels). See Appendix~\ref{envi_plot} for environment plots of other hosts. 

We also consider the environment of our hosts using the tidal index described in \cite{Karachen13} (see Tables~\ref{table:hosts_SMC}). The tidal index is a measure of the local stellar density derived from the stellar mass and distance of the nearest significant neighbors. $\Theta_5$  accounts for the tidal contributions from the five most significant neighbors \citep{Karachen13}.
The majority of our hosts have low tidal index values, with 22 hosts having $\Theta_5\leq0$, indicating they are isolated galaxies. The other 15 have higher tidal indices, indicating that they are not isolated galaxies but instead occupy a range of field and group environments. The tidal indices mostly agree with environment plots in Appendix~\ref{envi_plot}, as hosts with no known neighbors have $\Theta_5\leq0$, while those that appear to be in denser environments have $\Theta_5\geq0$. A few notable exceptions are discussed in the appendix.

\subsection{DESI Legacy Surveys Imaging Data} \label{data}

We utilize the publicly available, wide-field DESI Legacy Imaging Surveys  {dr-10} data \citep{Dey19} for satellite candidate detection.  For each host galaxy, we downloaded the $g$-band data from the web server\footnote{https://www.legacysurvey.org/} covering a projected radius of 150 kpc in small parts (12\arcmin$\times$12\arcmin~regions) and reconstructed larger fields with the astropy \citep{astropy13,astropy18} package \textit{reproject}. We preferred the larger fields of view to run the dwarf detection algorithm more efficiently.  { The exact \textit{g-}band depth varies across the survey area with an approximate 5$\sigma$ depth of \textit{g}$=$24.0.}  The data were retrieved with the native 0.262\arcsec~per pixel resolution and the ls-dr10 setting which merges the northern (MzLS$+$BASS) and southern (DECam) imaging data at the declination of 32.375\textdegree. 

The coverage for each host extends to a projected radius of 150 kpc.  This radius is slightly larger than the estimated 120$-$140~kpc virial radius for LMC-mass galaxies, and it extends well beyond the estimated 80$-$100~kpc virial radius of SMC-mass galaxies \citep{BMP2021}.

\section{Dwarf Satellite Search}\label{detect}

\subsection{ {Algorithm Search}}

\begin{figure*}[!th]
    \centering
    \includegraphics[width=\linewidth]{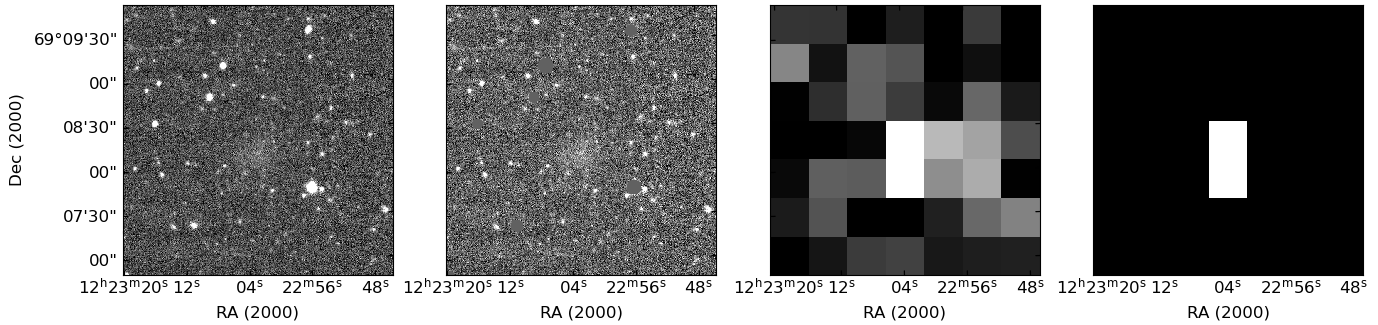}
    \caption{A demonstration of our detection algorithm, illustrating the steps to detect a dwarf candidate around NGC~4236 in the $g$-band of the DESI Legacy Survey imaging. Panels are 3\arcmin$\times$3\arcmin. Left panel: The original image of the dwarf candidate. Second panel: The image after masking objects from the GSC. Third panel: The image after spatially binning the masked image by 100$\times$100 pixels. Right panel: The final Source Extractor detected objects with $>$4~$\sigma$ above the background in the binned image. Detected objects at this stage are visually screened to remove clear false positives before being included in a private visual inspection gallery hosted on Zooniverse.} 
    \label{fig:algorithm}
\end{figure*}

We use a modified version of the  dwarf detection algorithm described in \cite{Bennet17} to identify candidate satellite galaxies.  This algorithm is based on previous works (e.g., \citealt{Dalcanton97, Davies16}).  The steps of the algorithm are illustrated in Figure~\ref{fig:algorithm} and briefly summarized below. 

The algorithm first masks foreground stars and known background galaxies in the Guide Star Catalog (GSC)~2.3.2 \citep{GSC} by creating a circular region covering the source and its outer halo.  The mask region grows in size logarithmically based on the magnitude of the source. This initial step is similar to the masking procedure used in other low-surface brightness galaxy searches (e.g., \citealt{vanderburg16, Bennet17,ELVES}). After bright source masking, Source Extractor \citep{Sextractor} is run on the masked image to identify sources with more than 25 pixels $>$5~$\sigma$ above the sky level. These relatively bright sources, such as the centers of high surface brightness galaxies, are masked without attempting to remove their outer halos.  The masked image is then spatially binned.  Two versions of the binned image are created, one binned by 100$\times$100 pixels and the other by 50$\times$50 pixels.  The binning corresponds to a spatial scale of $\sim$1200$\times$1200~pc ($\sim$600$\times$600~pc) at 10~Mpc and $\sim$500$\times$500~pc ($\sim$250$\times$250~pc) at 4~Mpc.  We choose these binning scales to detect large diffuse galaxies while remaining sensitive to smaller objects across the full distance range of the sample. We test different combinations of binning scales to optimize sensitivity and minimize false positives.  As seen in Figure~\ref{fig:algorithm}, our chosen binning scales enhances the detection of diffuse objects while effectively smoothing background variations. For both binning scales, pixels above 200$\sigma$ are masked to remove image artifacts such as chip gaps and diffraction spikes. Finally, we run the Source Extractor on the binned images, and pixels $\geq$4$\sigma$ above the background are cataloged.  The two catalogs from the different binning scales are compared, and any objects found in both catalogs are forwarded for initial visual screening.

 {To assess the survey completeness, we inject artificial galaxies into the data for each host galaxy using the Sersic2D function in \textit{astropy}. These artificial galaxies are simulated with a \sersic index of $n=1$ \citep{sersic68}, typical for diffuse dwarf galaxies \citep{koda15,vandokkum15,vanderburg16}. The simulated galaxies vary in absolute magnitudes from M$_g=$$-$7 to $-$13 (m$_g=$14$-$22.5 mag) and have effective radii (r$_e$) ranging from 2.6\arcsec\ to 185\arcsec\ (r$_e$=50$-$8720 pc).  The artificial galaxies have a uniform distribution in apparent magnitude from m$_g=$16.5$-$22.5 mag, with the number of artificial galaxies injected tapering off for m$_g\leq$16.5. The artificial galaxies have a logarithmic distribution in effective radius with more smaller galaxies injected. The number of artificial dwarfs injected into each bin per host is available at Zenodo \dataset[doi:10.5281/zenodo.15498887]{https://doi.org/10.5281/zenodo.15498887} and Github\footnote{https://github.com/hunte22l/ID-MAGE\_Completeness.git}.  For our detection efficiency tests, we model the artificial dwarfs as circular with zero ellipticity. As reported in \cite{vanderburg16}, moderate ellipticities do not impact the detection efficiency  {of the algorithm}. }

 {Artificial dwarfs are randomly injected with a uniformly weighted distribution across each image in batches.  To avoid affecting the Source Extractor's background measurement,  {fewer} large artificial dwarfs are injected in smaller batches. The total number of artificial galaxies injected per host galaxy is $\simeq$100,000$-$200,000, with the number of injected dwarfs increasing with the host's projected virial radius.  Following their injection into the DESI Legacy Survey data, we run our detection algorithm outlined above without the visual screening.}

 {The top panel of Figure~\ref{fig:full_comp} shows the detection efficiency for the entire survey in terms of apparent $g$ magnitude and $r_e$. Results for individual hosts are shown in Appendix~\ref{id_comp}. 
The completeness for individual hosts varies depending on image quality and depth but remains generally consistent across different hosts. Tables detailing the recovery rates per bin per host in terms of apparent magnitude and $r_e$ are available on Github, along with the number of artificial galaxies injected per magnitude and size bin.}

 {As depicted in Figure~\ref{fig:full_comp}, the algorithm's detection efficiency is $\gtrsim90\%$ for larger ($r_e\geq$5.5\arcsec), brighter ($m_g\leq$20.2) and higher central surface brightness objects ($\mu_{0,g}\lesssim$26.0$-$26.5), with the efficiency rapidly falling to below 50\% for galaxies with $\mu_{0,g}>26.5-27.0$ or $m_g>20.6-21.0$ for a $n=1$ \sersic profile, depending on the host. The algorithm also loses sensitivity to very compact, bright objects, such as bright elliptical galaxies and stars, as they are masked out. The algorithm completeness drops off between 20.0$\lesssim$$\mu_{0,g}$$\lesssim$21.0.  Based on \cite{Bennet17} and \cite{ELVES}, we expect very few dwarf galaxies to lie within this region. }

\subsection{ {Visual Screening}}

This screening is conducted through a web interface that presents the cutouts of the original $g$-band image, along with its masked, binned, and smoothed versions for each detected object. This display facilitates the easy identification of diffuse candidates and the removal of obvious false detections. 
To verify the effectiveness of our screening in identifying diffuse objects, we calibrate the display parameters using catalogs of known low-surface-brightness galaxies (SMUDGes; Systematically Measuring Ultra Diffuse Galaxies, \citealt{SMUDGESV}), diffuse satellite galaxies \citep{Bennet17,ELVES}, and simulated galaxies injected into our science fields. These simulated galaxies are modeled using the \textit{astropy.modeling} function Sersic2D as simple \sersic profiles covering the full range of  {expected} dwarf properties (m$_g=$16$-$22 mags, effective radius r$_e \simeq$ 2.6\arcsec$-$200\arcsec, \sersic index $n=$0.5$-$2.5, ellipticity of 0$-$0.4), and added before the masking process. We adjust the display settings for the screening to ensure consistent recovery of known low-surface-brightness galaxies and simulated diffuse galaxies. Simulated galaxies fail screening if they are obscured by galactic cirrus or superimposed on stars or other galaxies.

In a typical field of view, $\sim$70$-$100 objects per square degree are forwarded for visual screening.  {The number of objects which pass visual screening per square degree varies depending on the distance of the host, with $\sim$2$-$3 objects passing for the nearest hosts and $\sim$4$-$6 objects passing for the more distant ones. Overall, about one in 17 objects passed the visual screening for further visual inspection.} Most false detections are background galaxies, galaxy clusters, or unmasked halos around bright stars and galaxies. 

In total, ID-MAGE covers $\sim$224 square degrees, including $\sim$168 square degrees around SMC-mass hosts and $\sim$56 square degrees around LMC-mass hosts. A total of  {$\sim$1180} detections pass the visual screening for further visual inspection.

\subsection{Visual Inspection} \label{zoo}

After the initial screening, we conduct a systematic visual inspection of our candidate satellites in a private Zooniverse project, using $g$, $r$, and $z$-band three-color images. On Zooniverse, experts evaluate image cutouts by answering the prompt: \textit{Identify this object (is this a satellite galaxy?)} with three response options: \textit{satellite galaxy}, \textit{massive or distant galaxy}, and \textit{not a galaxy/image defect}. Each object is classified by at least six team members to establish a final score.
During the visual inspection, objects that do not display distinct structures, such as spiral arms or a bulge, are rated as likely satellite galaxies. 
Additionally, objects that are either diffuse and low surface brightness or blue and clumpy are rated highly.
This meticulous visual inspection aims to eliminate contaminants, such as probable background galaxies, and to prioritize follow-up observations for the most promising satellite candidates.

In Zooniverse, we include $\sim$110 galaxies with confirmed distances or velocities that are detected by the algorithm and pass the initial screening. We use responses to these known objects to calibrate the scoring system for our candidate satellite sample. Notably, 85\% of known galaxies that receive unanimous scores as potential satellites fall within the distance range (4$-$10~Mpc) and velocity range (v$\leq$1200 km s$^{-1}$) of our host galaxies. The lowest score for a known galaxy within the expected velocity/distance range is two-thirds agreement on its classification as a potential satellite galaxy. Consequently, galaxies receiving unanimous agreement as potential satellites are considered excellent candidates and are grouped into the high-likelihood sample. Those with at least two-thirds agreement constitute the full ID-MAGE sample of candidate satellites. Additionally, any candidate with a known distance or velocity in the literature confirming it to be a satellite is included in the high-likelihood sample rather than the full sample.

Objects scoring below two-thirds agreement are considered false positives and are not presented here.  {Approximately 65\% of the candidates that pass the initial screening also pass the Zooniverse visual inspection. The number of objects which pass each step are presented in Table~\ref{table:Candidates}.   Overall, we identify one candidate satellite for approximately every $\sim30$ objects detected by the algorithm, aligning with findings from \cite{Bennet17} and similar searches (e.g., \citealt{Vollmer13,2Merrit14,vanderburg16,ELVES}).}    Table~\ref{Table:4449} showcases our list of candidates, detailing each candidate's proposed host, ratings, photometric properties, and whether it is confirmed as a satellite based on a distance/velocity measurement in the literature.

\begin{table} 
    \centering
    \small
    \caption{\centering Number of Candidates Identified per Step}
        \begin{tabular}{l l l l} 
        \hline
        \hline 
        Step & Per deg$^{2}$ & Total \\ [0.5ex] 
        \hline 
        Algorithm & 70-100 & $\sim$20000  \\
        Visual Screening & 5 & $\simeq$1180 \\
        Visual Inspection (Pass) & 3.4 & 763  \\
        Visual Inspection (Highly Rated) & 1.0 & 229 \\
        Morphology Cut (Full Sample) & 1.6 & \tot \\
        Morphology Cut (High Likelihood) & 0.6 & \high \\[1ex]
        \hline 
    \end{tabular} 
    \label{table:Candidates} 
\end{table}

\begin{figure*}

\includegraphics[width=\linewidth]{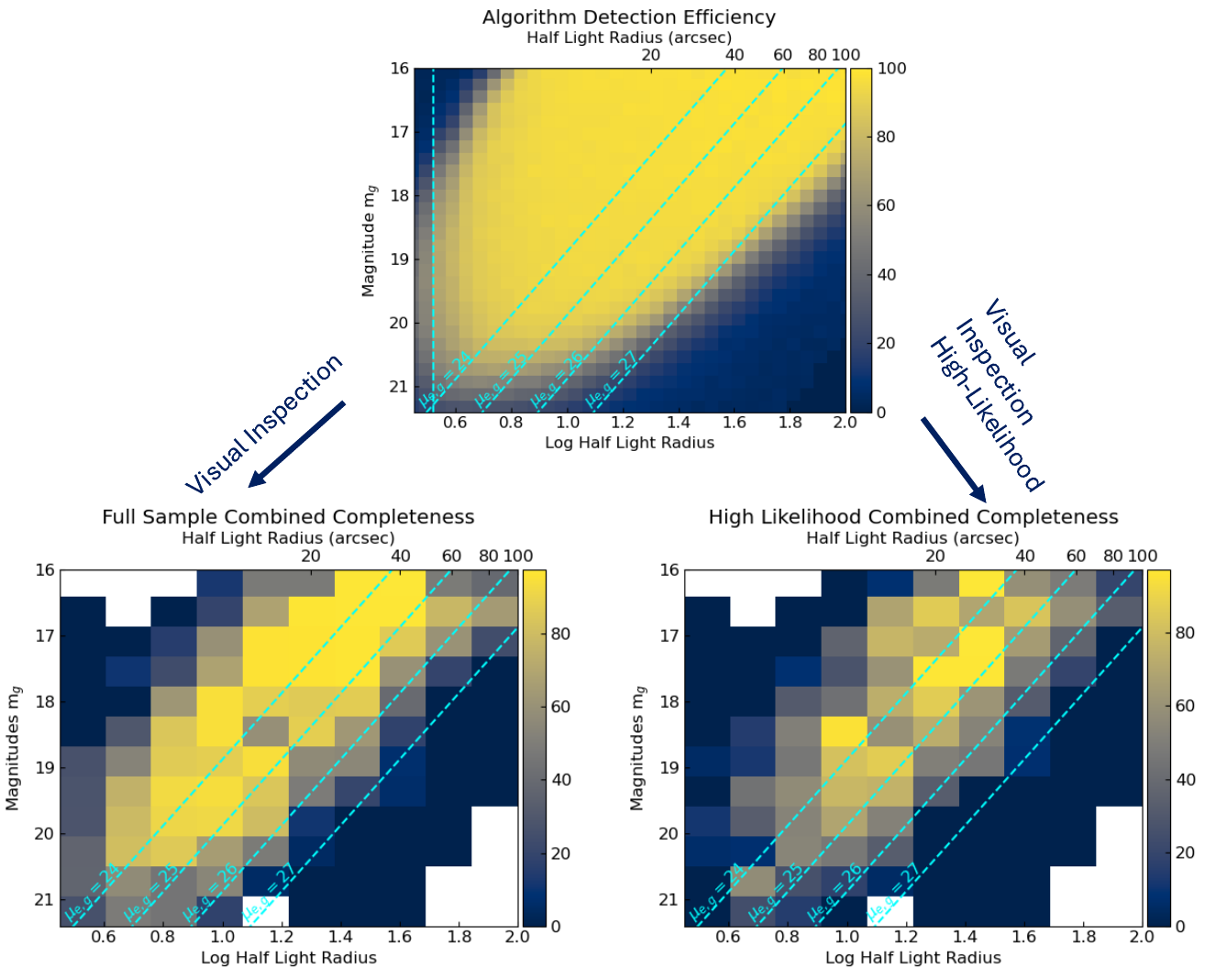}
\caption{Average completeness of our dwarf search as quantified by image simulations with injected artificial galaxies. Top panel: The algorithm detection efficiency in terms of apparent magnitude and size. Bottom-left panel: The overall completeness of our survey, which combines the algorithm detection efficiency and the Zooniverse visual inspection, for the full sample. {Bottom-right panel:} The combined completeness of our survey for the high-likelihood sample (candidates unanimously identified as satellite candidates during visual inspection). Overplotted in each figure are lines of constant central surface brightness ($\mu_{0,g}=$24$-$27~mag arcsec$^{-2}$) assuming a \sersic index of 1. There is a drop in completeness at low surface brightness ($\mu_{0,g}>26$~mag arcsec$^{-2}$), which is driven by a drop in galaxy identification during the visual inspection.  The algorithm is complete to a surface brightness $\sim$0.5 mag arcsec$^{-2}$ fainter than the combined completeness.}
\label{fig:full_comp}
\end{figure*}

To assess the impact of our visual inspection on completeness, we include $\sim$1,100 artificial dwarf galaxies in the Zooniverse visual inspection step. These fake galaxies are injected across all science fields to account for variations in depth and image quality. They possess the same range of properties as those used in the initial visual screening  {(m$_g=$16$-$22 mags, effective radius r$_e \simeq$ 2.6\arcsec$-$200\arcsec, \sersic index $n=$0.5$-$2.5, ellipticity of 0$-$0.4)}. Each artificial dwarf is assigned a $g-r$ color between 0 and 1.0, and a $g-z$ color between 0.1 and 1.5, matching the color ranges observed in galaxies from surveys like ELVES and SAGA. These artificial dwarfs receive high ratings on Zooniverse, with good completeness down to $\mu_g \sim$ 26 mag arcsec$^{-2}$.   {The visual inspection's ratings of the artificial dwarfs are not impacted by the morphology of the artificial dwarfs.  The artificial galaxies which passed visual inspection have the same distribution of ellipticities, \sersic indices, \textit{g-r} colors, and \textit{g-z} colors as the overall sample included in the Zooniverse sample. The visual inspection ratings are most strongly driven by surface brightness.  The visual inspection falls off for brighter, more compact objects with higher surface brightnesses ($\mu_g<$23 mag arcsec$^{-2}$)  }   In Figure~\ref{fig:full_comp}, the bottom panels show our recovery rates after visual inspection. The left panel represents the full sample, while the right panel focuses on the high-likelihood sample, which consists of galaxies unanimously identified as satellite candidates. We further validate our recovery rates for visual inspection by incorporating $\sim$80 diffuse ($\mu_{0,g}\gtrsim23.5$, $r_e>$10\arcsec) satellite galaxies from the ELVES survey. These ELVES satellites are recovered with the same completeness as the artificial dwarfs, confirming the effectiveness of our visual inspection procedures.

\subsection{Survey Completeness} \label{completeness}

To fully quantify the completeness of our sample, we multiple the visual inspection completeness with the algorithm's detection efficiency.  The combined completeness thresholds are primarily determined by the visual inspection limits at low surface brightnesses. Specifically, the visual inspection achieves completeness down to $\mu_{0,g} \sim$ 26 mag arcsec$^{-2}$, which is approximately 0.5~mag arcsec$^{-2}$ brighter than the algorithm's completeness limit of $\mu_{0,g}\sim$ 26.5 mag arcsec$^{-2}$. This highlights the importance of accounting for both algorithmic and visual completeness in surveys that rely on a visual inspection component. 

The completeness in absolute magnitude depends on distance and so varies with host (see Appendix~\ref{id_comp}). At the median distance of our host galaxies (6.25~Mpc), our 90\% completeness limit of $m_g\leq$20.2 is M$_g=-8.8$. We find that our survey is complete down to roughly M$_V\sim-9$, assuming $g-V\simeq$0.25.
Our recovery rates are somewhat lower than those reported by \cite{Bennet17}, and  similar to the ones observed in the ELVES and  {ELVES-Dwarf surveys}. This arises from the varying depths of the photometric data employed by these studies. Specifically, \cite{Bennet17} utilized deep imaging from the Canada-France-Hawaii Telescope Legacy Survey (CFHTLS), and ELVES used a combination of DESI Legacy Survey dr-9 data and archival CFHT/MegaCam data. Our data is all from the Legacy Survey  {dr-10} fields, and so is closest to the ELVES survey in terms of photometric completeness. 

In this paper, we focus on candidates that lie within the high completeness regime of our survey. Therefore, we only include candidates with $M_V\lesssim-9$ and $r_e>3.1$\arcsec (as the completeness drops below 50\% for objects smaller than 3.1\arcsec).  {Additionally, there are very few known galaxies with $M_V <-9$ that have an $r_e$ less than 100~pc (e.g., \citealt{Brasseur,ADD23,Pace24}).  We consider candidates with $r_e <100$~pc significantly more likely to be background galaxies rather than satellites.  Therefore, to remove likely interlopers from the sample, we impose an additional physical size criteria cut, assuming the candidate is at the distance of the presumed host.  To accommodate uncertainties in our photometric measurements (see Section~\ref{galfit}) and to avoid removing potential satellites, we include candidates with $1.15\times r_e >$100 pc.  This cut removes 22 objects from the sample, all of which are around hosts within 5.5~Mpc.   See Table~\ref{Table:4449} for the complete candidate sample organized by host and candidate quality.}

\subsection{Known Objects}

As part of our search process, we also check for known candidate satellite galaxies that the search algorithm may have missed. We identify five galaxies that are not recovered by the algorithm due to their close proximity to bright objects, resulting in partial or full masking. Three of these galaxies have known velocities or distances that align with their assumed hosts, while the other two do not have distance/velocity measures. We include these five galaxies in Table~\ref{Table:4449}. Additionally, our algorithm successfully identified many previously known galaxies, which are incorporated into the main sample upon passing visual inspection. Most of these recovered galaxies come from previous photometric searches for low surface brightness galaxies (LSBGs, e.g., SMUDGes; \citealt{SMUDGESV}, MATLAS \citealt{Duc15,Habas20}) and lack known distances or velocities. 
The algorithm detects 89\% of the SMUDGes LSBGs within the footprint of the survey with a \textit{g}$<$21.0. 
In total, we detect \tot candidates, with \new newly identified candidate satellites and 16 galaxies with published distance/velocities, confirming them as satellites.

\section{Properties of Candidate Satellites} \label{galfit}

\begin{deluxetable*}{l l l l l l l l l l l l l}
\centerwidetable
\tabletypesize{\scriptsize}
\tablecaption{ID-MAGE Candidate Satellites Properties from GALFIT Photometry \label{Table:4449}}
\tablehead{\colhead{ Name} & \colhead{RA} & \colhead{Dec} & \colhead{Host} & \colhead{$m_g$} & \colhead{$m_r$} & \colhead{A/B} & \colhead{$M_V$} & \colhead{$r_e$} & \colhead{$\mu_{0,g}$} & \colhead{$M_*$} & \colhead{\sersic} & \colhead{Set} \\
\colhead{  } & \colhead{J2000} & \colhead{J2000} & \colhead{ } & \colhead{mag} & \colhead{mag}  & \colhead{ } & \colhead{mag} & \colhead{arcsec} & \colhead{mag$^{\arcsec^{-2}}$} & \colhead{$\log(M_*)$} & \colhead{ } & \colhead{ } \\
\colhead{ (1)} & \colhead{(2)} & \colhead{(3)} & \colhead{(4)} & \colhead{(5)} & \colhead{(6)} & \colhead{(7)} & \colhead{(8)} & \colhead{(9)} & \colhead{(10)} & \colhead{(11)} & \colhead{(12)}  & \colhead{(13)} }
\startdata
MAGE J0245-6030 & 02h45m12.55s & -60d30m58.32s & ESO115-G21 & 17.4$\pm$0.1 & 17.2$\pm$0.1 & 0.42$\pm$0.01 & -11.2$\pm$0.3 & 7.1$\pm$0.1 & 21.8$\pm$0.1 & 6.3$\pm$0.2 & 0.93$\pm$0.1 & H \\
MAGE J0231-6205 & 02h31m08.18s & -62d05m16.44s & ESO115-G21 & 19.7$\pm$0.1 & 19.3$\pm$0.1 & 0.68$\pm$0.01 & -9.0$\pm$0.3 & 4.7$\pm$0.1 & 23.3$\pm$0.1 & 5.7$\pm$0.2 & 1.09$\pm$0.1 & F \\
MAGE J0243-6111 & 02h43m20.33s & -61d11m38.04s & ESO115-G21 & 19.3$\pm$0.1 & 18.9$\pm$0.1 & 0.47$\pm$0.01 & -9.4$\pm$0.3 & 4.5$\pm$0.1 & 22.6$\pm$0.1 & 5.9$\pm$0.2 & 1.0$\pm$0.1 & F \\
MAGE J0233-6125 & 02h33m04.68s & -61d25m34.68s & ESO115-G21 & 18.5$\pm$0.1 & 18.5$\pm$0.1 & 0.55$\pm$0.01 & -10.0$\pm$0.3 & 6.8$\pm$0.1 & 22.5$\pm$0.1 & 5.7$\pm$0.2 & 1.23$\pm$0.1 & H \\
MAGE J0252-5526 & 02h52m13.66s & -55d26m58.92s & ESO154-G23 & 18.9$\pm$0.1 & 18.5$\pm$0.1 & 0.29$\pm$0.01 & -10.1$\pm$0.3 & 6.0$\pm$0.1 & 21.6$\pm$0.1 & 6.1$\pm$0.2 & 1.39$\pm$0.1 & F \\
MAGE J0248-5403 & 02h48m38.18s & -54d03m44.64s & ESO154-G23 & 19.7$\pm$0.1 & 19.3$\pm$0.1 & 0.59$\pm$0.01 & -9.4$\pm$0.3 & 5.1$\pm$0.1 & 23.0$\pm$0.1 & 5.8$\pm$0.2 & 1.26$\pm$0.1 & F \\
MAGE J0246-5456 & 02h46m45.10s & -54d56m01.68s & ESO154-G23 & 18.4$\pm$0.1 & 18.2$\pm$0.1 & 0.6$\pm$0.01 & -10.5$\pm$0.3 & 4.9$\pm$0.1 & 22.9$\pm$0.1 & 6.0$\pm$0.2 & 0.52$\pm$0.1 & H \\
MAGE J0304-5335 & 03h04m01.27s & -53d35m05.64s & ESO154-G23 & 19.4$\pm$0.1 & 19.2$\pm$0.1 & 0.73$\pm$0.01 & -9.5$\pm$0.3 & 3.2$\pm$0.1 & 22.9$\pm$0.1 & 5.6$\pm$0.2 & 0.7$\pm$0.1 & F \\
MAGE J0139-4510 & 01h39m51.86s & -45d10m27.84s & ESO245-G05 & 19.3$\pm$0.1 & 19.1$\pm$0.1 & 0.4$\pm$0.01 & -9.1$\pm$0.3 & 5.1$\pm$0.1 & 23.5$\pm$0.1 & 5.5$\pm$0.2 & 0.48$\pm$0.1 & H \\
MAGE J0144-4314 & 01h44m08.14s & -43d14m01.32s & ESO245-G05 & 18.5$\pm$0.1 & 18.3$\pm$0.1 & 0.58$\pm$0.01 & -9.8$\pm$0.3 & 5.7$\pm$0.1 & 23.5$\pm$0.1 & 5.7$\pm$0.2 & 0.39$\pm$0.1 & H \\
MAGE J0143-4324 & 01h43m29.23s & -43d24m34.92s & ESO245-G05 & 18.5$\pm$0.1 & 18.1$\pm$0.1 & 0.64$\pm$0.01 & -9.9$\pm$0.3 & 10.1$\pm$0.1 & 24.1$\pm$0.1 & 6.0$\pm$0.2 & 0.88$\pm$0.1 & H \\
MAGE J0140-4221 & 01h40m57.41s & -42d21m05.40s & ESO245-G05 & 19.6$\pm$0.1 & 18.9$\pm$0.1 & 0.48$\pm$0.01 & -9.1$\pm$0.3 & 5.8$\pm$0.1 & 23.1$\pm$0.1 & 6.0$\pm$0.2 & 1.21$\pm$0.1 & F \\
MAGE J0142-4316 & 01h42m08.59s & -43d16m43.68s & ESO245-G05 & 18.5$\pm$0.1 & 18.2$\pm$0.1 & 0.56$\pm$0.01 & -9.9$\pm$0.3 & 4.4$\pm$0.1 & 22.3$\pm$0.1 & 6.0$\pm$0.2 & 0.81$\pm$0.1 & F \\
MAGE J0143-4413 & 01h43m51.86s & -44d13m48.36s & ESO245-G05 & 19.2$\pm$0.1 & 18.4$\pm$0.1 & 0.36$\pm$0.01 & -9.5$\pm$0.3 & 5.9$\pm$0.1 & 21.8$\pm$0.1 & 6.4$\pm$0.2 & 1.6$\pm$0.1 & F \\
MAGE J0606-3246 & 06h05m59.78s & -32d46m38.64s & ESO364-G29 & 20.7$\pm$0.1 & 20.0$\pm$0.1 & 0.8$\pm$0.01 & -9.4$\pm$0.3 & 5.6$\pm$0.5 & 24.2$\pm$0.1 & 6.1$\pm$0.2 & 1.45$\pm$0.12 & H \\
MAGE J0606-3252 & 06h06m40.37s & -32d52m26.04s & ESO364-G29 & 19.1$\pm$0.1 & 18.7$\pm$0.1 & 0.84$\pm$0.01 & -10.9$\pm$0.3 & 4.0$\pm$0.1 & 22.8$\pm$0.1 & 6.3$\pm$0.2 & 0.99$\pm$0.1 & F \\
MAGE J0608-3247 & 06h08m44.93s & -32d47m04.56s & ESO364-G29 & 18.1$\pm$0.1 & 17.7$\pm$0.1 & 0.63$\pm$0.01 & -11.9$\pm$0.3 & 3.8$\pm$0.1 & 21.9$\pm$0.1 & 6.8$\pm$0.2 & 0.69$\pm$0.1 & F \\
MAGE J0605-3249 & 06h05m47.54s & -32d49m45.12s & ESO364-G29 & 19.4$\pm$0.1 & 18.7$\pm$0.1 & 0.47$\pm$0.01 & -10.8$\pm$0.3 & 10.5$\pm$0.6 & 23.6$\pm$0.1 & 6.8$\pm$0.2 & 1.55$\pm$0.1 & H \\
MAGE J0605-3220 & 06h05m46.30s & -32d20m48.84s & ESO364-G29 & 18.8$\pm$0.1 & 18.6$\pm$0.1 & 0.77$\pm$0.01 & -11.0$\pm$0.3 & 3.6$\pm$0.1 & 21.4$\pm$0.1 & 6.3$\pm$0.2 & 1.44$\pm$0.1 & F \\
MAGE J0606-3257 & 06h06m16.39s & -32d57m51.48s & ESO364-G29 & 17.9$\pm$0.1 & 17.6$\pm$0.1 & 0.61$\pm$0.01 & -12.0$\pm$0.3 & 5.0$\pm$0.1 & 22.2$\pm$0.1 & 6.8$\pm$0.2 & 0.74$\pm$0.1 & H \\
MAGE J0609-3225 & 06h09m09.91s & -32d25m22.08s & ESO364-G29 & 17.0$\pm$0.1 & 16.8$\pm$0.1 & 0.65$\pm$0.01 & -12.9$\pm$0.3 & 11.1$\pm$0.1 & 23.1$\pm$0.1 & 6.8$\pm$0.2 & 0.69$\pm$0.1 & H \\
MAGE J0601-3306 & 06h01m52.34s & -33d05m59.64s & ESO364-G29 & 20.5$\pm$0.1 & 19.7$\pm$0.1 & 0.53$\pm$0.01 & -9.7$\pm$0.3 & 13.3$\pm$1.6 & 26.9$\pm$0.1 & 6.4$\pm$0.2 & 0.74$\pm$0.1 & F \\
\enddata
\tablecomments{Column~1: ID-MAGE identifier. Column~2: the Right Ascension (J2000.0). Column~3: the Declination (J2000.0). Column~4: the presumed host of the candidate. Column~5: Apparent $g-$band  magnitude. Column~6: Apparent $r-band$ magnitude. Column~7: A/B Axial ratio. Column~8: Absolute $V-band$ magnitude. Column~9: Effective radius in arcsec. Column~10: Central surface brightness in the $g-$band in mag arcsec$^{-2}$. Column~11: \sersic index measured with GALFIT. Column~12: Derived stellar mass from \textit{g-} and \textit{r-}band magnitudes. Column~13: category that candidate belongs to: 
C=Confirmed, H=High-likelihood, F=Full Sample. \\
See electronic version for the full table. }
\end{deluxetable*}

We use GALFIT \citep{Peng_2010} to measure the structural parameters of our candidates. We implement a fitting approach akin to the first fitting stage in \cite{khim+24}, outlined as follows. 

We fit a single \sersic profile to each galaxy in the $g$-band, avoiding any potential influence from stellar clumps or overlapping objects. To achieve this, we not only mask nearby objects, but also mask any central region containing a minimum of 5 adjacent pixels with a surface brightness 1.5 times (0.44 mag) brighter than 24 mag arcsec$^{-2}$. We utilize the PSF provided by the DESI Legacy Survey and calculate pixel-by-pixel uncertainties ($\sigma$-images) using the inverse-variance images from the DESI Legacy Survey. We adopt a flat background to prevent over-subtraction of the galaxy's wings. The convolution box size for GALFIT is configured to 26.2\arcsec$\times$26.2\arcsec\ (100$\times$100 pixels), which is half the image length of 52.4\arcsec$\times$52.4\arcsec\ (200$\times$200 pixels). The free parameters are the following: central position, \sersic index, $r_e$, magnitude, axis ratio, position angle, and background level. 

To minimize the influence of our initial parameter selections, we repeat the fitting procedure six times using various initial estimations. We employ combinations of two different effective radii (30 and 50 pixels) and three surface brightness values at the effective radii (20, 25, 30 mag arcsec$^{-2}$). We calculate the reduced chi-squared statistic, $\chi^2_\nu$, within a circular region of radius 50 pixels (approximately 13.1\arcsec) centered on the image center and opt for the model with the lowest $\chi^2_\nu$ value. Occasionally, GALFIT yields a model fit with huge final parameter uncertainties. We only regard models where $r_e > 2\sigma_{r_e}$ as meaningful.  For the \textit{r}-band, we use the effective radius, central position, \sersic index, axis ratio, and position angle from the \textit{g}-band fit and only fit for magnitude and background level.  The GALFIT results for each candidate are compiled in Table~\ref{Table:4449}.   {The errors reported here are based on the uncertainties returned by GALFIT. We adopt a minimum uncertainty of 0.1 when GALFIT reports a smaller value, as it significantly underestimates the true photometric errors (see \citealt{Bennet17,SMUDGESV}).  We will determine and report the full errors when we finalize our sample after our follow-up campaign (see Section~\ref{follow-up}).}

We convert our \textit{g} and \textit{r}-band photometry to  {V-band using the Lupton (2005)\footnote{http://www.sdss3.org/dr8/algorithms/sdssUBVRITransform.php} transformation for the Sloan Digital Sky Survey:}
\begin{equation}
    V=g-0.5784(g-r)-0.0038
\end{equation}

To estimate the stellar mass of our galaxies,  {we use the color-corrected stellar mass calibration for low-mass galaxies published in \cite{dlReyes24}:
\begin{equation}
    log(M_*/M_\odot) = 1.433(g-r) + 0.00153M^2_{g,0} - 0.335 M_{g,0} + 2.072
\end{equation}}

\section{The Satellite Candidates} \label{discussion}

Around our 36 host galaxies, we identify \tot satellite candidates in the full sample and \high in the high-likelihood sample. This section provides an overview of our candidates and the satellite systems.  We first summarize the number of satellites identified per host. In Section~\ref{sec_comp}, we compare the photometric properties of our satellites with those from the ELVES survey and dwarf galaxies within $\sim$3~Mpc. In Section~\ref{contaminant}, we assess the possible contamination rate of our sample and use it to  {estimate the lower range }of our satellite luminosity function.  Finally, Section~\ref{sec_lf} discusses  {the upper range of} our sample and the luminosity function derived from the high-likelihood candidates.

We detect \highlm high-likelihood candidates (candidates with unanimous agreement as potential satellites from the visual inspection) around the 9 LMC-mass hosts. The richest LMC-mass satellite systems are IC~1727/NGC~0672, hosting 10 high-likelihood candidates between two hosts, and NGC~4244 with 6 high-likelihood candidates. In contrast, the least satellite-rich LMC-mass systems are NGC~0024 and NGC~6503, each with only 2 high-likelihood candidates. The mean number of high-likelihood satellite candidates per system LMC-mass hosts is $\sim4$. 

We identify \highsm high-likelihood candidates around the 27 SMC-mass hosts. The richest SMC-mass satellite system is NGC~3738, which has 15 high-likelihood candidates. However, this high number appears to be driven by a significant number of background galaxies contaminating the system (see Sections \ref{contaminant}). Following NGC~3738, the next richest SMC-mass system is IC~4182, with 8 high-likelihood candidates. The system with the fewest candidates is UGC~04426, which has no high-likelihood candidates. Including NGC~3738, the mean number of high-likelihood satellite candidates per SMC-mass host is $\sim$3.5, but excluding NGC~3738, the mean drops to $\sim$3 satellites per system.

\subsection{Comparison with Known Local Volume Dwarf Galaxies} \label{sec_comp}

\begin{figure*}[!th]
    \centering
    \includegraphics[width=\linewidth]{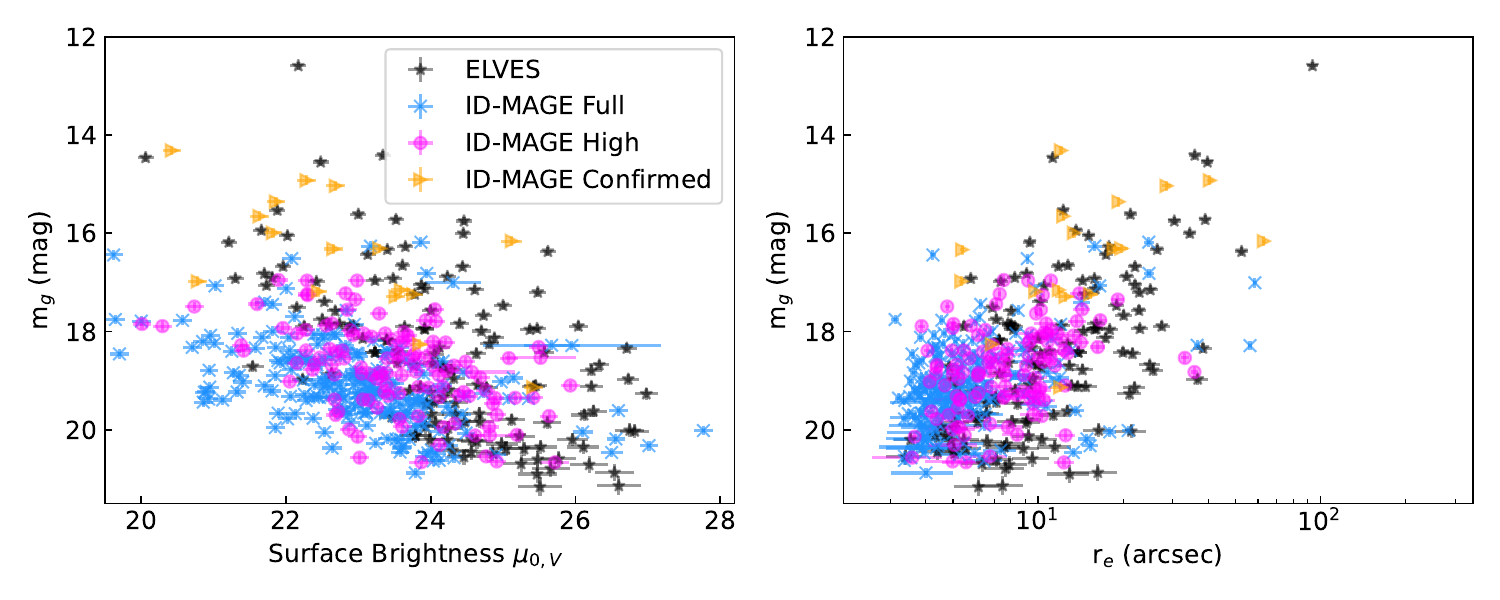}
    \caption{A comparison of the properties of  {our confirmed satellites (orange triangle)}, our high-likelihood candidates (magenta circles), the remaining candidates in our full satellite sample (blue crosses), and satellites of MW-mass hosts (ELVES; black stars). The left panel shows V-band central surface brightness versus apparent g-band (m$_g$), while the right panel presents effective radius ($r_e$ in arcseconds) versus m$_g$.  The ID-MAGE satellite candidates exhibit a similar range in photometric properties to the known satellites around MW-mass hosts.}
    \label{fig:ELVES_full}
\end{figure*}

We compare the GALFIT photometry of the satellite candidates with known satellite galaxies, with the ELVES survey sample being the most comparable due to its similar distance range, though their hosts are MW-mass. As shown in Figure~\ref{fig:ELVES_full}, the two samples align well in apparent $g$-band magnitude, surface brightness, and $r_e$ in arcseconds. 
The ELVES sample includes brighter satellites compared to our survey, which is expected due to our survey's focus on lower-mass hosts.  The most massive ELVES satellites are comparable in mass to our hosts. Within our full sample, there is a distinct group of candidates that are compact and have higher surface brightness than the ELVES satellites.   {These candidates are located in lower-left in both panels of Figure~\ref{fig:ELVES_full}. Based on their compact sizes ($r_e\sim$3.5\arcsec), faint magnitudes ($m_g\sim18-20$), and higher surface brightnesses ($\mu_{0,g}\lesssim24.0$), these objects are likely background galaxies. } ELVES performed follow-up surface brightness fluctuation (SBF) distance measurements on similar galaxies and found that the majority were background galaxies. Follow-up distance measurements will likely remove many of these candidates from our sample.

\begin{figure}[!t]
    \centering
    \includegraphics[width=\linewidth]{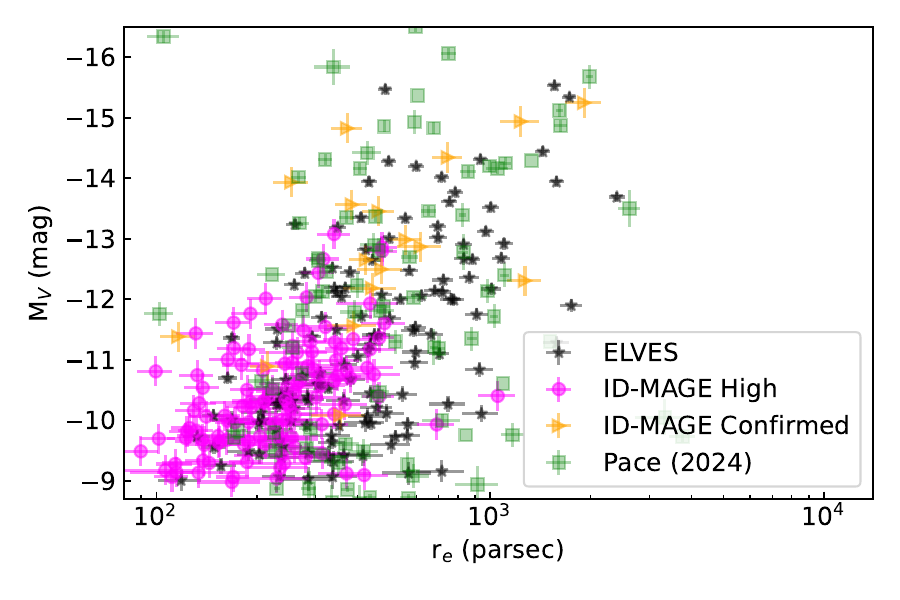}
    \caption{A comparison of the effective radius (in parsecs) and absolute V-band magnitude (M$_V$) of our high-likelihood candidate sample (magenta circles, assuming the distance of the presumed host), our confirmed satellites (yellow triangles), satellites of MW-mass hosts (ELVES; black stars), and dwarf galaxies within $\sim$3~Mpc from \cite{Pace24} (green squares). 
    } 
    \label{fig:ELVES_high}
\end{figure}

Figure~\ref{fig:ELVES_high} compares the ID-MAGE high-likelihood sample to the ELVES sample and dwarf galaxies within $\sim$3~Mpc \citep{Pace24}, demonstrating that their derived physical properties are generally similar. However, our sample contains fewer galaxies with $r_e\geq$500~pc compared to both the ELVES sample and the \citet{Pace24} compilation.   
As shown in Figure~\ref{fig:full_comp}, this difference is not due to the completeness of our survey, which is $\simeq$60$-$90\% for the full sample and $\simeq$50–70\% for the high-likelihood sample in this region. It is predicted that satellites of LMC/SMC-mass hosts are less massive and fainter than the satellites of MW-mass hosts.  Given the size-luminosity relation, it is not surprising that we do not identify many satellites with $r_e\geq$500~pc.  Additionally, as illustrated in Figure~\ref{fig:ELVES_high}, most high-likelihood candidates with $r_e\geq$500~pc are confirmed satellites, indicating that there are very few potential false positives in this region, unlike the less massive candidates where few have published distances/velocities.    

\subsection{Sample Contamination} \label{contaminant}

Contamination is a well-documented challenge for unresolved satellite galaxy searches (e.g., \citealt{Bennet19,Bennet20,ELVES}). In some cases, like NGC~3738 and NGC~4244, likely background galaxy groups can be identified within the candidate sample. These hosts exhibit an unusually high concentration of satellite candidates in small regions of their search area. 
Many candidates around NGC~3738 are likely associated with NGC~3718, a Seyfert 1 galaxy at 14 Mpc, or a galaxy cluster at 21~Mpc including NGC~3733 and NGC~3756, based on their close proximity to the massive galaxies and each other. Due to its potentially higher contamination rate, we exclude NGC~3738 from further analysis to prevent skewing our sample results. 

\begin{figure*}[!t]
    \centering
    \includegraphics[width=\linewidth]{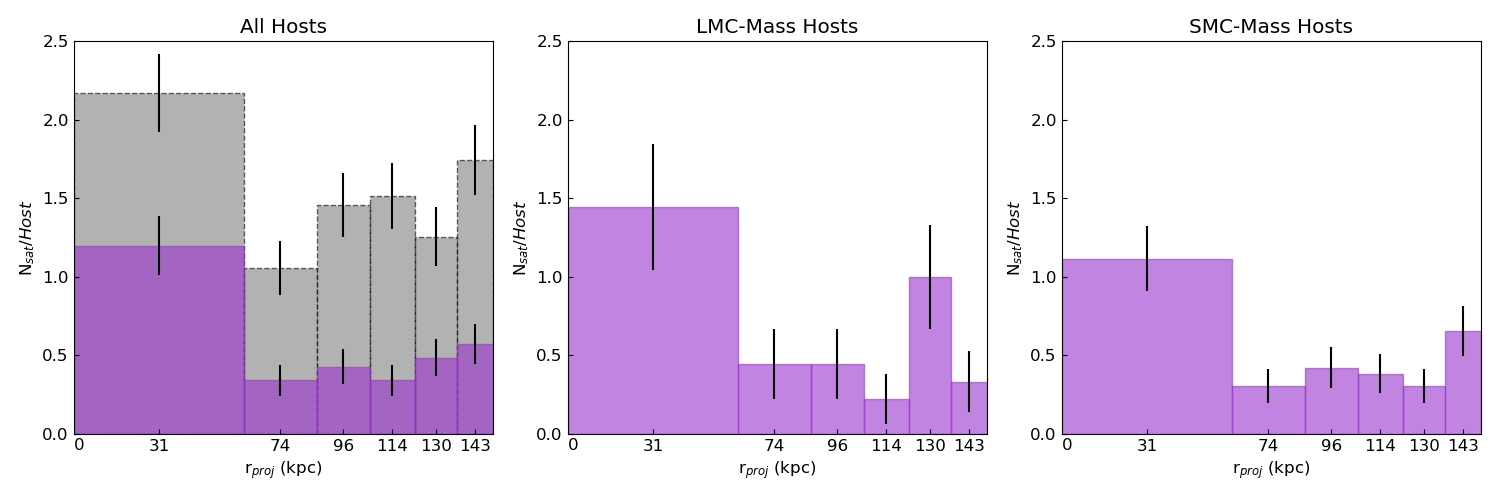}
    \caption{Central concentration of our candidate satellite sample: the number of candidates per equal-area annular bin per host as a function of projected radius (excluding NGC~3738). Left panel: The full sample (gray) and the high-likelihood sample (purple). Center panel: The high-likelihood candidates around LMC-mass hosts. Right: The high-likelihood candidates around SMC-mass hosts. Error bars represent the Poisson uncertainties per bin. The projected distance to each satellite is based on the known distance of its assumed host.  { The satellite candidates are spatially binned into equal-area annular bins.}  This equal-area binning highlights the concentration of candidate satellites within $\sim60$~kpc of their hosts.  For the full sample, there is a minimum of one candidate per host per bin. In the center and right panels, the number of candidates decreases in outer bins due to fewer contaminants in our high-likelihood samples.  A central concentration of our candidate satellites is evident for both LMC-mass and SMC-mass hosts. } 
    \label{fig:dist_host}
\end{figure*}

While the sample certainly includes some false positives, there is strong evidence that a significant portion are real satellites. By spatially binning the satellite sample into equal-area annular bins, we observe a distinct concentration of candidates within 60 kpc of their host, compared to the outer bins (see Figure~\ref{fig:dist_host}).  {The number of satellites per bin should correlate with the distance from the host. However, the number of background galaxies should be roughly constant as the number should only depend on area and the annular rings are equal in area.}  {This assumption of a uniform surface density of the background contamination has been used and tested previously for imaging based satellite searches around galaxies with masses of 9.5$\leq$log(M$_*$/M$_\odot$)$\leq$11 (e.g., \citealt{xSAGA}).}  { However, the background distribution for LMC-mass systems may not be as uniform compared higher-mass systems.} The central concentration seen in Figure~\ref{fig:dist_host} aligns with the models from \cite{Dooley17a,Dooley17}, which are based on the \textit{Caterpillar} simulation suite \citep{Griffin16}. They predict that approximately 50–65\% of satellites around an LMC-mass galaxy should lie within 50–60 kpc of their host. This central concentration is even more pronounced in the high-likelihood sample, suggesting it has a lower contamination rate.  {The ELVES and SAGA satellite samples demonstrate a clear central concentration in Figure 9 of \cite{ELVES} which compares the number of satellites per host in equal width annular bins. For the satellites in the ELVES sample, the central concentration is strongest for the more massive satellites, which is not seen for the SAGA satellite sample. }

The number of background contaminants in each bin should be roughly equal. To estimate the number of background galaxies per bin, we use the outer bins in Figure~\ref{fig:dist_host}. For the high-likelihood sample, the average number of contaminants per host per bin in the outer five bins for our LMC-mass hosts is 0.49$\pm$0.10 for LMC-mass hosts and 0.42$\pm$0.06 for our SMC-mass hosts. The number for the innermost bin is 1.4$\pm$0.40 for LMC-mass hosts and 1.12$\pm$0.21 for SMC-mass hosts. After accounting for the average number of contaminants from the outer bins, the innermost bin shows an excess of 0.91$\pm$0.41 candidates for LMC-mass hosts and 0.70$\pm$0.22 candidates for SMC-mass hosts. If we assume that approximately 60\% of our satellites reside within 60 kpc of their host \citep{Dooley17}, this provides a {lower estimate} of \lowlimlm (\lowlimsm) identified satellites per LMC-mass (SMC-mass) host in the high-likelihood sample. This aligns well with the lower range of cosmological predictions from \cite{Dooley17a} (see Section~\ref{sec_lf}),  { and the lower estimate for our LMC-mass hosts is in agreement with the satellite luminosity function of NGC~2403 (an LMC-mass host) in \cite{Carlin24}.}

\subsection{Satellite Luminosity Function} \label{sec_lf}

Satellite surveys have found a correlation between the satellite abundance and stellar mass of the host for MW-mass galaxies \citep{SAGAII, Carlsten21a, Danieli23,BMP24}. Hydrodynamical simulations using different abundance matching models and stellar-mass to halo-mass relations ($M_{\star}$$-$$M_{halo}$) broadly predict that as the host's halo mass increases, the stellar mass increases and the number of satellites within a given mass range also increases (e.g., \citealt{Santos-Santos22}).  \cite{ELVES} found that the simulated luminosity functions from the ARTEMIS simulations \citep{Font2021} agree well with the average satellite abundance of MW-mass hosts at higher mass. However, the predictions may modestly underestimate the satellite abundance of the less massive hosts. This demonstrates a need to constrain the satellite abundances of galaxies less massive than the MW. Observations of the satellite luminosity functions (LF) of dwarf galaxies provide vital tests to different abundance-matching models and constraints to the slope of the $M_*-M_{halo}$ relationship. 

\begin{figure}
    \centering
    \includegraphics[width=\linewidth]{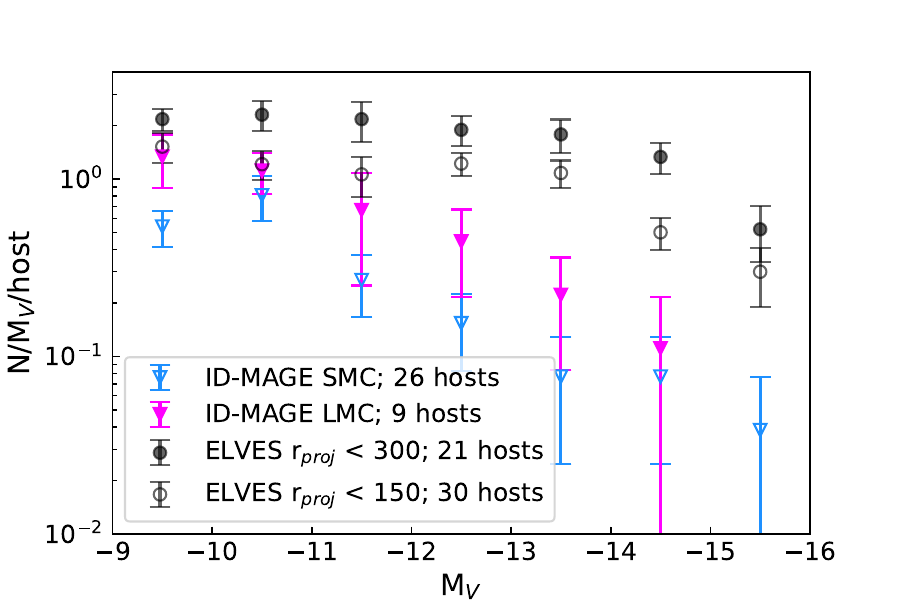}
    \caption{ {The upper estimate of the satellite luminosity function of our hosts}, using the high-likelihood sample (excluding NGC~3738). Figure shows the average satellite abundance per host in 1 magnitude wide bins within the assumed virial radius. ELVES survey sample is shown in black.   
    The legend indicates the number of hosts contributing to the stacked bins. As expected, the number of satellites per mag bin increases as the host stellar mass increases. 
    } 
    \label{fig:host_to_host}
\end{figure}

 {To estimate an upper range of our hosts' LF, we consider a maximum number of satellites we detect. We assume that the visual inspection is entirely reliable and that all high-likelihood candidates will be confirmed as satellites.  We combine the high-likelihood sample with the already confirmed satellites as our upper estimate.}  As discussed in Section~\ref{contaminant}, our high-likelihood sample consists of the candidates with lower contamination from background and foreground galaxies. Additionally, for SMC-mass hosts, we apply a virial radius cutoff of 110~kpc —slightly larger than the estimated virial radius of the SMC \citep{BMP2021}—to exclude probable interlopers as the area searched extends to 150~kpc. Figure~\ref{fig:host_to_host} presents the LF for our hosts with these assumptions. To account for variations in satellite abundance between hosts, the error bars are the standard deviation of the number of satellites per magnitude bin per host, divided by $\sqrt{N_{host}}$.  {Figure~\ref{fig:host_to_host} also shows the LF based on our lower estimates of the number of satellites per host from the central concentration plots.}

For the LMC and SMC-mass hosts, the LF 
appears to rise as the absolute magnitude becomes fainter.  Overall, LMC-mass hosts tend to have more satellites per magnitude bin than SMC-mass hosts.  This is consistent with simulations \citep{Dooley17,Dooley17a,Santos-Santos22} which predict LMC-mass galaxies host richer satellite systems than SMC-mass galaxies. This difference is not due to the larger virial radius used for LMC-mass hosts. Expanding the SMC-mass host radius to 150~kpc slightly increases the number of satellite candidates per bin, but the overall trend remains unchanged. 

Figure~\ref{fig:host_to_host} illustrates that our hosts have significantly fewer satellites per magnitude bin than the MW-mass hosts of the ELVES survey. 
Comparing the three host mass ranges, there is a clear trend that as the host mass increases, so does the number of satellites in each magnitude bin. This trend aligns well with predictions from the $\Lambda$CDM model that satellite abundance correlates with host mass (e.g., \citealt{Dooley17a,Dooley17,Santos-Santos22}).  This also agrees with the correlation \cite{Carlsten21a} observed between the stellar masses of MW-mass hosts and their satellite abundances. Our sample extends this observed trend to the dwarf host galaxy mass range. 

\begin{figure*}
    \centering
    \includegraphics[width=\linewidth]{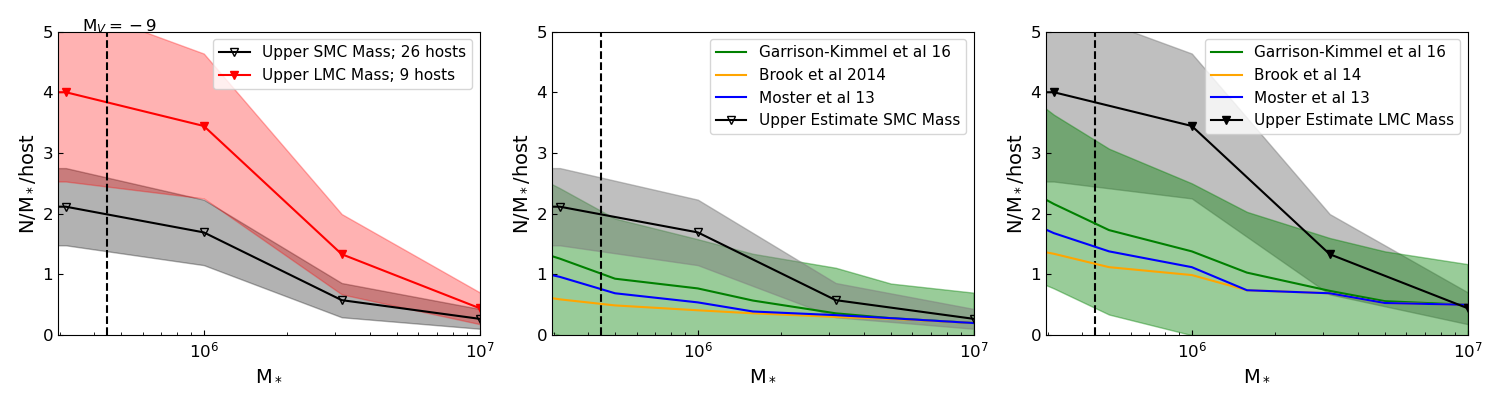}

    \caption{ {The upper estimate of the satellite LF for LMC/SMC-mass hosts. Left panel: The average satellite abundance per host (excluding NGC~3738) as a function of minimum stellar mass within the assumed virial radius for our hosts.} The legend indicates the number of hosts contributing to each LF.  The LMC-mass hosts consistently have more satellites per mass bin than the SMC-mass hosts. Center/Right panel: Satellite LF of our SMC/LMC-mass hosts, in comparison to theoretical predictions. 
    The candidate satellites are shown in black with the gray shaded region representing the standard deviation of the number of satellites per magnitude bin per host, divided by $\sqrt{N_{host}}$. 
    The theoretical predictions are taken from \cite{Dooley17}, which are based on 
    the dark matter simulation suite \textit{Caterpillar} \citep{Griffin16}, combined with different stellar-mass halo-mass ratios \citep{Moster13,Brook14,GK16}. The shaded green region shows the 1$\sigma$ variation in satellite abundance from \citeauthor{GK16} model. } 
    \label{fig:LF}
\end{figure*}

In Figure~\ref{fig:LF}, we compare our LF with theoretical predictions taken from \cite{Dooley17}.  These predictions are derived from the dark matter-only \textit{Caterpillar} simulation suite \citep{Griffin16}, combined with different $M_*$-$M_{halo}$ relations \citep{Moster13, Brook14, GK16}.  Since most of our satellites are unconfirmed, our LF represents an upper limit of the true luminosity function. On average, LMC-mass hosts contain \LMCLFer satellites per host, while SMC-mass hosts have \SMCLFer satellites per host with M$_V<$-$9$. The shaded region shows the standard deviation of the number of satellites per magnitude bin per host, divided by $\sqrt{N_{host}}$.  

The upper estimate of our LF trends toward the higher end of predictions, particularly for the LMC-mass hosts.  The upper-estimate LF falls within the 1$\sigma$\ range of predictions based on the \cite{GK16} $M_*-M_{halo}$ relation.   The more massive satellites--where a larger fraction are confirmed --show good agreement with predictions for both LMC- and SMC-mass hosts.  {Further, our upper estimate for the LF of the SMC-mass hosts is in agreement with the LF of NGC 3109 (a SMC-mass host) in \cite{ADD3109}}  {Our lower limits of \lowlimlm (\lowlimsm) candidates per LMC-mass (SMC-mass) host with M$_V<$-$9$ are similar to the number satellite identified and confirmed per host by ELVES-Dwarf \citep{ELVES-Dwarf}. \cite{ELVES-Dwarf} confirmed zero to two satellites per SMC-mass host and zero to one satellites per LMC-mass host.}

 {Our upper-estimate is an over-estimate; however, more than half of the high-likelihood candidates would need to be interlopers for the final confirmed sample to be below the lower range of the predictions in \cite{Dooley17}. }   {As seen in Figure~\ref{fig:LF},} our lower estimates from Section~\ref{contaminant}, \lowlimsm (SMC-mass) and \lowlimlm (LMC-mass), align with the lower-range of predictions from \cite{Dooley17} and \cite{Dooley17a}.  Therefore, the number of confirmed satellites after our follow-up observations is expected to fall within these bounds.  {Additionally, the true shape of the LF is likely different due to the interlopers in our sample and because we have not applied a completeness correction. The flattening of the LF for the lowest mass bin is likely due to the lower completeness in that bin, which can be attributed in part to the lower detection completeness for our more distant hosts. } Once finalized, our sample will enable a more thorough comparison between observations and theoretical predictions, offering the first statistically significant observational constraints on the number of satellites around LMC- and SMC-mass hosts. 

\section{Follow-Up Campaign} \label{follow-up}

The ultimate goal of ID-MAGE is to create the {\it first statistical} view of dwarf satellites around low-mass hosts, substantially extending the range of host masses and environments probed by existing surveys (see Figure~\ref{fig:host_scatter}), delivering quantitative constraints for galaxy formation physics in $\Lambda$CDM. We are currently conducting a comprehensive observational follow-up campaign to confirm and characterize our satellite candidates. As a first step, we check our candidates against the \hi~surveys HIPASS \citep{HIPASS} and ALFALFA \citep{ALFALFA} because if we can identify \hi~line emission, we can immediately confirm or refute the association with the host.  For those without \hi~redshifts, we are leading two complementary follow-up campaigns to measure their distances: (1) deeper imaging follow-up for the SBF technique, and (2) optical spectroscopic follow-up for redshift measurements. This will allow us to confidently identify and remove interlopers, resulting in a clean catalog of satellites around our host galaxies. To characterize our satellites, we are leading an \hi~observation campaign, coupled with a GALEX UV photometric study, to study their gas content and star formation activity. Additionally, we will compare satellite and system properties (e.g., star formation rates, quenched fraction, gas fraction) with recent studies investigating the environmental impact of MW-host on dwarf galaxies  \citep[e.g.,][]{Karunakaran20,Karunakaran21,Karaunkaran22,Karunakaran23,Jones24}. Our follow-up campaign will enable us to examine the gas retention and star formation activity in individual satellites, as well as the quenched fraction and luminosity function across the satellite systems.

While visual inspection effectively reduces false positives, contamination from background galaxies often remains significant, and can account for a large proportion of satellite candidates (e.g., \citealt{ELVES,Bennet19,Bennet20}).  The contamination rate can be as high as $\sim$80\% in searches around MW-mass galaxies in cases where there is a massive galaxy in the background with its own satellite system. Thus, follow-up observations are essential to verify that these candidates are genuine satellites of the purported hosts. Our follow-up campaign is actively underway, with several current and recently completed observational runs. 

SBF offers an efficient way to measure distances to quenched dwarfs (i.e., ones without star-forming regions) in the distance range of our hosts \citep{Carlsten19,carlsten19a}. Although the DESI Legacy Imaging Surveys provide full virial volume coverage of our host galaxies, which is essential for a systemic dwarf search, the data are too shallow and the seeing is too poor to perform SBF distance measurements. Therefore, we are leading a deep follow-up imaging campaign with MMT, Gemini North, and Magellan to use the SBF technique efficiently. 

It is well known that the SBF technique is not ideal for gas-rich, star-forming systems because their star-forming regions can significantly affect the SBF measurements \citep{Greco21}. Therefore, for our blue candidates with patchy morphology, we are simultaneously undergoing a ground-based spectroscopic campaign to obtain redshift measurements. Similar campaigns have successfully obtained redshifts of diffuse dwarf galaxies from clear emission \citep[e.g.,][]{Greco2018} or absorption lines \citep[e.g.,][]{vanDokkum2015,Kadowaki2017}. As part of this campaign, we have secured observational time to conduct long-slit spectroscopy using several telescopes, including Gemini/GMOS, LBT/MODS, Magellan/IMACS, MMT/Binospec, SALT/RSS, and MDM/OSMOS.

For some candidates, confirming their status as satellite galaxies requires both distance and velocity measurements. As illustrated in Appendix~\ref{envi_plot}, the virial radii of some hosts overlap with those of other galaxies that are also massive enough to host their own satellite systems. For candidates in these overlapping regions, distance or velocity alone is insufficient to confidently associate a satellite with its host. Our comprehensive follow-up strategy is designed to address this challenge, 
and we will publish the results of our ongoing follow-up campaign in a future papers.

\hi~is the initial fuel for star formation, so its presence or lack thereof in satellites enables a better understanding of their past and future evolution. We have been awarded $\sim$200 hours on the Green Bank Telescope (GBT; PI: Hunter) to assess whether our candidates are gas-poor, old stellar systems, or gas-rich, recently star-forming dwarfs. Our observational strategy is designed to give a near uniform sensitivity across all targets in terms of their \hi-to-stellar mass ratio, with our targets being $M_{HI}/M_{\star}$$=$$1$ (at $5\sigma$). Given that $M_{HI}/M_{\star}$$>$$3$ for gas-rich field dwarfs \citep{Huang12}, even for undetected targets, we will be able to confidently conclude that they are gas-poor. 
Furthermore, we will compare the \hi measurements with star formation rates derived from archival GALEX UV data. This combined data set will enable us to determine the quenched fraction of satellites around low-mass hosts and explore the mechanisms responsible for quenching.  

\section{Conclusions} \label{conclusion}

This paper presents the first overview of our new survey, ID-MAGE, which is designed to identify satellites of 36 low-mass host galaxies with distances between 3.9 and 10~Mpc. Our survey aims to analyze the characteristics of individual satellites (e.g., morphology and scaling relations) and the properties of the satellite systems (e.g., dwarf galaxy abundance, luminosity function, and quenched fraction).  To achieve this, we analyze DESI Legacy Survey imaging data of the area around 9 LMC-mass and 27 SMC-mass hosts out to a distance of 150~kpc.  

We employ the detection algorithm described in \cite{Bennet17} with additional visual inspection by experts. 
We assess the completeness of our algorithm using artificial dwarf injections. Additionally, we evaluate the completeness of our visual inspection. Combined, the algorithm and visual inspection are complete down to $M_V\lesssim -9$ and $\mu_{0,g}\simeq26$ mag arcsec$^{-2}$. 

In total, we identify \tot satellite candidates, including \new newly discovered galaxies. Among these, \high are classified as high-likelihood candidates based on our systematic visual inspection. Of the \high high-likelihood candidates, \highlm are associated with LMC-mass hosts and \highsm with SMC-mass hosts. The number of satellite candidates per hosts ranges from 0 to 15 high-likelihood candidates.  This scatter may be driven by physical factors, such as environmental richness or intrinsic host-to-host scatter. However, some of this variation is likely due to foreground and background contaminants. The candidates in our sample range in apparent magnitudes between 15.0~$<m_g<$~20.9 and effective radii between 3.1\arcsec and 75\arcsec. The photometric properties of our sample are consistent with satellites of MW-mass galaxies and dwarf galaxies within 3~Mpc \citep{Pace24}.  

For our hosts, we identify fewer satellites per magnitude bin per host compared to MW-mass hosts and we find fewer candidates per magnitude bin for the SMC-mass hosts compared to the LMC-mass hosts. This agrees with $\Lambda$CDM simulations that predict as the host's halo mass increases, the stellar mass increases, and so to does the number of satellites (e.g. \citealt{Dooley17,Dooley17a,Santos-Santos22}). Our low-mass hosts also exhibit the trend observed among MW-mass hosts, where satellite abundance correlates with host stellar mass, extending this relationship into the dwarf host galaxy mass range.

From our candidate sample, we establish {upper and lower estimates of the LF for low-mass galaxies.  To determine the  {lower estimate}, we analyze the central concentration of candidates around their hosts, estimating a rough lower bound of \lowlimlm (\lowlimsm) satellites per LMC-mass (SMC-mass) host with M$_V\leq-$9. The high-likelihood sample serves as the upper estimate, with \LMCLFer (\SMCLFer) candidates per LMC-mass (SMC-mass) host. Our upper and lower estimates bracket the predicted range for satellites of dwarf galaxies from \cite{Dooley17, Dooley17a}. 

We are currently conducting deep imaging and spectroscopic follow-up campaigns to confirm and characterize our satellite candidates.  To efficiently follow-up the \tot candidates, we are utilizing a large range of facilities, such as training telescopes, GBT L-band observations, and poor weather time.  Additionally, we have secured observational time to follow-up our targets photometrically and spectroscopically with telescopes around the world, including GBT, Gemini/GMOS, LBT/MODS, Magellan/IMACS, MMT/Binospec, SALT/RSS, and MDM/OSMOS.  We will publish the results of our ongoing follow-up campaign in future papers.

Our survey, ID-MAGE, already provides valuable insight into how host mass influences satellite populations. Moving forward, follow-up observations will refine our catalog, enabling detailed analysis of how host mass and environment affect satellite populations. 
This will lead to a deeper understanding of the galaxy formation and evolution processes in the $\Lambda$CDM  paradigm.
 
\begin{acknowledgments}


DJS and the Arizona team acknowledges support from NSF grant AST-2205863.

KS acknowledges funding from the Natural Sciences and Engineering Research Council of Canada (NSERC).

We thank the anonymous reviewer for their comments and suggestion which have significantly improved the quality of this paper.

This research has made use of the NASA/IPAC Extragalactic Database (NED), which is operated by the Jet Propulsion Laboratory, California Institute of Technology, under contract with NASA.

The Legacy Surveys consist of three individual and complementary projects: the Dark Energy Camera Legacy Survey (DECaLS; Proposal ID \#2014B-0404; PIs: David Schlegel and Arjun Dey), the Beijing-Arizona Sky Survey (BASS; NOAO Prop. ID \#2015A-0801; PIs: Zhou Xu and Xiaohui Fan), and the Mayall z-band Legacy Survey (MzLS; Prop. ID \#2016A-0453; PI: Arjun Dey). DECaLS, BASS and MzLS together include data obtained, respectively, at the Blanco telescope, Cerro Tololo Inter-American Observatory, NSF’s NOIRLab; the Bok telescope, Steward Observatory, University of Arizona; and the Mayall telescope, Kitt Peak National Observatory, NOIRLab. Pipeline processing and analyses of the data were supported by NOIRLab and the Lawrence Berkeley National Laboratory (LBNL). The Legacy Surveys project is honored to be permitted to conduct astronomical research on Iolkam Du’ag (Kitt Peak), a mountain with particular significance to the Tohono O’odham Nation.

NOIRLab is operated by the Association of Universities for Research in Astronomy (AURA) under a cooperative agreement with the National Science Foundation. LBNL is managed by the Regents of the University of California under contract to the U.S. Department of Energy.

This project used data obtained with the Dark Energy Camera (DECam), which was constructed by the Dark Energy Survey (DES) collaboration. Funding for the DES Projects has been provided by the U.S. Department of Energy, the U.S. National Science Foundation, the Ministry of Science and Education of Spain, the Science and Technology Facilities Council of the United Kingdom, the Higher Education Funding Council for England, the National Center for Supercomputing Applications at the University of Illinois at Urbana-Champaign, the Kavli Institute of Cosmological Physics at the University of Chicago, Center for Cosmology and Astro-Particle Physics at the Ohio State University, the Mitchell Institute for Fundamental Physics and Astronomy at Texas A\&M University, Financiadora de Estudos e Projetos, Fundacao Carlos Chagas Filho de Amparo, Financiadora de Estudos e Projetos, Fundacao Carlos Chagas Filho de Amparo a Pesquisa do Estado do Rio de Janeiro, Conselho Nacional de Desenvolvimento Cientifico e Tecnologico and the Ministerio da Ciencia, Tecnologia e Inovacao, the Deutsche Forschungsgemeinschaft and the Collaborating Institutions in the Dark Energy Survey. The Collaborating Institutions are Argonne National Laboratory, the University of California at Santa Cruz, the University of Cambridge, Centro de Investigaciones Energeticas, Medioambientales y Tecnologicas-Madrid, the University of Chicago, University College London, the DES-Brazil Consortium, the University of Edinburgh, the Eidgenossische Technische Hochschule (ETH) Zurich, Fermi National Accelerator Laboratory, the University of Illinois at Urbana-Champaign, the Institut de Ciencies de l’Espai (IEEC/CSIC), the Institut de Fisica d’Altes Energies, Lawrence Berkeley National Laboratory, the Ludwig Maximilians Universitat Munchen and the associated Excellence Cluster Universe, the University of Michigan, NSF’s NOIRLab, the University of Nottingham, the Ohio State University, the University of Pennsylvania, the University of Portsmouth, SLAC National Accelerator Laboratory, Stanford University, the University of Sussex, and Texas A\&M University.

BASS is a key project of the Telescope Access Program (TAP), which has been funded by the National Astronomical Observatories of China, the Chinese Academy of Sciences (the Strategic Priority Research Program “The Emergence of Cosmological Structures” Grant \# XDB09000000), and the Special Fund for Astronomy from the Ministry of Finance. The BASS is also supported by the External Cooperation Program of Chinese Academy of Sciences (Grant \# 114A11KYSB20160057), and Chinese National Natural Science Foundation (Grant \# 12120101003, \# 11433005).

The Legacy Survey team makes use of data products from the Near-Earth Object Wide-field Infrared Survey Explorer (NEOWISE), which is a project of the Jet Propulsion Laboratory/California Institute of Technology. NEOWISE is funded by the National Aeronautics and Space Administration.

The Legacy Surveys imaging of the DESI footprint is supported by the Director, Office of Science, Office of High Energy Physics of the U.S. Department of Energy under Contract No. DE-AC02-05CH1123, by the National Energy Research Scientific Computing Center, a DOE Office of Science User Facility under the same contract; and by the U.S. National Science Foundation, Division of Astronomical Sciences under Contract No. AST-0950945 to NOAO.

This publication uses data generated via the Zooniverse.org platform, development of which is funded by generous support, including a Global Impact Award from Google, and by a grant from the Alfred P. Sloan Foundation.

\textit{Facilities}: Blanco - Cerro Tololo Inter-American Observatory's 4 meter Blanco Telescope, Mayall - Kitt Peak National Observatory's 4 meter Mayall Telescope, Hiltner - Michigan-Dartmouth-MIT Observatory 2.4 meter Telescope

\textit{Software:} Astropy \citep{astropy13,astropy18}, GALFIT \citep{Peng_2010}, Matplotlib \cite{Hunter:2007}, NumPy \citep{harris2020array}, pandas \citep{mckinney-proc-scipy-2010}, Source Extractor \citep{Sextractor}.

\end{acknowledgments}

\appendix

\section{Environment Plots} \label{envi_plot}

This section provides details on the environment of specific host galaxies.   {The online figure set for Figure 2 illustrates} the environment of each host along with its candidate satellites. The surveyed areas are shaded gray, and prominent galaxies 
are displayed with their approximate virial radii: 110~kpc for SMC-mass galaxies, 150~kpc for LMC-mass galaxies, and 300~kpc for more massive galaxies. In some cases, host search areas overlap, causing a few candidates to be found in multiple hosts' search areas.  If a candidate is found in more than one searched area, we associate it with the physically closest host, assuming it lies at the same distance.  Follow-up observations (see Section \ref{follow-up}) will confirm the true hosts of these candidates.

\noindent
\textbf{NGC~0625:} NGC~0625 (3.92~Mpc) and ESO245-G05 (4.46~Mpc) have overlapping virial radii; however, they are not considered a single system like IC~1727 and NGC~0672 due to differences in their distances. 

\noindent
\textbf{NGC~4244:} NGC~4244 (4.20~Mpc) appears to be part of a group of LMC-mass and smaller galaxies, including NGC~4395(4.65~Mpc)).  This group of low-mass galaxies has measured TRGB distances between 4 and 5~Mpc \citep{EDD}. The galaxies CGCG187-05, UGC~7559, and UGC~7599 lie just beyond NGC~4244's virial radius, with TRGB distances of 4.85 Mpc, 4.97 Mpc, and 4.72~Mpc, respectively \citep{EDD}. These distances place them more than 500 kpc behind NGC~4244, suggesting they belong to the same galaxy group but are unlikely to be its satellites.

\noindent
\textbf{NGC~4449:} NGC~4449's (4.16~Mpc) virial radius overlaps with that of M94, which is at a similar distance (4.3~Mpc; \citealt{EDD}), indicating that NGC~4449 and its satellites are likely members of the larger M94 galaxy group.

\noindent
\textbf{IC~4182:} IC~4182's (4.24~Mpc) virial radius overlaps with that of M94, which is at a similar distance (4.3~Mpc \citealt{EDD}), indicating that IC~4182 and its satellites are likely members of the larger M94 galaxy group. 

\noindent
\textbf{NGC~4236:}  NGC~4236's (4.31~Mpc) virial radius overlaps with that of UGC~07490; however, the two are unlikely to be associated due to their large difference in velocities. In Cosmicflows-4, NGC~4236 has a measured velocity of $\simeq$0 km s$^{-1}$ \citep{EDD}, while UGC~07490 has a velocity of 468$\pm$5 km s$^{-1}$.

\noindent
\textbf{ESO245-G05:} NGC~0625 (3.92~Mpc) and ESO245-G05 (4.46~Mpc) have overlapping virial radii; however, they are not considered a single system like IC~1727 and NGC~0672 due to differences in their distances.

\noindent
\textbf{NGC~4395:} NGC~4395 (4.65~Mpc) appears to be part of a group of LMC-mass and smaller galaxies, including NGC~4244 (4.20~Mpc). This group has measured TRGB distances between 4 and 5~Mpc \citep{EDD}. UGC~07698 and UGC~07605 lie just beyond NGC~4395's estimated virial radius and appear to belong to the same galaxy group based on their Cosmicflows-4 distances.

\noindent
\textbf{NGC~5585:} NGC~5585 (6.84~Mpc)) is located on the outskirts of a galaxy group.  Its virial radius overlaps that of M101, which is at a similar distance (6.7 Mpc; \citealt{EDD}).  

\noindent
\textbf{IC~1727/NGC~0672:} NGC~0672 (7.0~Mpc) and IC~1727 (7.29~Mpc) have high tidal indices due to their close proximity to each other.  


\noindent
\textbf{UGC~04115:} UGC~04115 (7.7~Mpc) appears relatively isolated; however, UGC~03974, a SMC-mass galaxy with three times the mass of UGC~04115, is located at nearly the same distance (7.99~Mpc) and has a projected distance of 500 kpc.  This places UGC~03974 just outside the plotted area, suggesting this may be a small group of low-mass galaxies. 

\noindent
\textbf{UGC~03974:} UGC~03974 (7.99~Mpc) appears relatively isolated; however, UGC~04115, a SMC-mass galaxy with one-third the mass of UGC~03974, is located at nearly the same distance (7.7~Mpc) and has a projected distance of 500 kpc.  This places UGC~04115 just outside the plotted area, suggesting this may be a small group of low-mass galaxies. 

\noindent
\textbf{NGC~2188:} NGC~2188 (8.22~Mpc), ESO364-G29 (8.81~Mpc), and HIPASSJ0607-34 (9.4~Mpc) have overlapping virial radii; however, they are not considered a single system like IC~1727 and NGC~0672 due to the differences in their distances. 

\noindent
\textbf{ESO364-G29:} NGC~2188 (8.22~Mpc), ESO364-G29 (8.81~Mpc), and HIPASSJ0607-34 (9.4~Mpc)  have overlapping virial radii; however, they are not considered a single system like IC~1727 and NGC~0672 due to the differences in their distances.

\noindent
\textbf{NGC~3432:} NGC~3432 (8.9~Mpc) appears very isolated, but has a high tidal index of $\Theta_5=3.3$. This high tidal index is partially due to UGC~05983 (8.9~Mpc), a relatively massive satellite ($M_*\simeq4$$\times$$10^7$) very close to NGC~3432. Excluding UGC~05983 from the tidal index calculation, results in $\Theta_5=2.7$.

\noindent
\textbf{HIPASSJ0607-34:} NGC~2188 (8.22~Mpc), ESO364-G29 (8.81~Mpc), and HIPASSJ0607-34 (9.4~Mpc),  have overlapping virial radii; however, they are not considered a single system like IC~1727 and NGC~0672 due to the differences in their distances.

\noindent
\textbf{ESO486-G21:} For ESO486-G21 (9.7~Mpc), the nearby galaxy NGC~1744 does not have a TRGB distance measurement but has a measured velocity of 741 km s$^{-1}$ \citep{springbob05}.  This is similar to ESO486-G21's velocity of 840 km s$^{-1}$ \citep{HIPASS}. Additionally, multiple Tully-Fischer distance measurements place NGC~1744 within 1~Mpc of ESO486-G21 (e.g. \citealt{EDD}), suggesting that these two galaxies may be associated and part of the same group.

\section{Individual Host Completeness Results} \label{id_comp}

This section presents the overall completeness of each host, incorporating both algorithmic detection efficiency and visual inspection completeness  {(see the online figure set for Figure 4)}, along with details on specific hosts and their satellite candidates. These completeness plots also show the candidate satellites for each host. However, galaxies with m$_g<$16.0 are excluded, as they are previously known satellites, not new discoveries from our survey. 

\noindent
\textbf{NGC~4707: UGC~07950, a known satellite of NGC~4707, is not shown in Figure~\ref{fig:full_comp}.  It has a SBF distance of 7.04~Mpc from ELVES-Dwarf and a projected separation of 56~kpc from NGC~4707.  UGC~07950 has an estimated stellar mass of log(M$_*$/M$_\odot$)$\simeq$7.4$\pm0.3$ and is approximately a fourth of the mass of NGC~4707. }

\noindent
\textbf{UGC~03974:} KK98~65, a known satellite of UGC~03974, is not shown in \textbf{Figure~\ref{fig:full_comp}}.  It has a TRGB distance of 7.98~Mpc from Cosmicflows-4 and a projected separation of 39~kpc from UGC~03974.  KK98~65 has an estimated stellar mass of log(M$_*$/M$_\odot$)$\simeq$7.3$\pm0.6$ and is approximately a tenth of the mass of UGC~03974. 

\noindent
\textbf{NGC~3432:} UGC~05983, a known satellite of NGC~343, is not shown in \textbf{Figure~\ref{fig:full_comp}}.  It has a projected separation of 8.2~kpc from NGC~3432  and a relative velocity of 80 km s$^{-1}$. With an estimated stellar mass of log(M$_*$/M$_\odot$)$\simeq$7.6$\pm0.6$, UGC~5983 is one of the most massive satellites in the sample, approximately 1/30 of the mass of NGC~3432.  It is one of the known galaxies in our sample that the algorithm did not recover due to masking.

\noindent
\textbf{NGC~4861:} CVnIIdwA, a likely satellite of NGC~4861, is not shown in \textbf{\ref{fig:full_comp}}. It has a projected distance of 100 kpc from NGC~4861 and a relative velocity of 103 km s$^{-1}$.  CVnIIdwA is one of the most massive satellites in sample with an estimated stellar mass of log(M$_*$/M$_\odot$)$\simeq$7.8$\pm0.6$, approximately one-tenth of the mass of NGC~4861.  Another confirmed satellite of  NGC~4861, KK98~175 is also not shown. KK98~175 has a projected distance of 100 kpc and a relative velocity of 130 km s$^{-1}$. Both satellites are located at the outer edge of NGC~4861's estimated virial radius of $\sim100$~kpc.

{\small\bibliography{ref}}

\end{document}